\documentclass[aps,twocolumn,pra, 10pt, floatfix, longbibliography, final]{revtex4-1}
\usepackage{graphicx, bm, amsmath, amsfonts, amssymb, color}
\usepackage[caption=false]{subfig}

\usepackage[%
  pdftex,
  breaklinks=true,
  colorlinks=true,
  linkcolor=blue,
  urlcolor=blue,
  citecolor=blue
]{hyperref}

\begin{document}

\title{Impact of nonlocal electrodynamics on the flux noise and inductance of superconducting wires}
\author{Pramodh Senarath Yapa}
\author{Tyler Makaro}
\author{Rog\'erio de Sousa}
\affiliation{Department of Physics and Astronomy, University of Victoria, Victoria, British Columbia, Canada V8W 2Y2}

\date{\today}

\begin{abstract}
We present exact numerical calculations of supercurrent density, inductance, and impurity-induced flux noise of cylindrical superconducting wires in the nonlocal Pippard regime, which occurs when the Pippard coherence length is larger than the London penetration depth. In this regime the supercurrent density displays a peak away from the surface, and changes sign inside the superconductor, signalling a breakdown of the usual approximation of local London electrodynamics with a renormalized penetration depth. Our calculations show that the internal inductance and the bulk flux noise power is enhanced in nonlocal superconductors. In contrast, the kinetic inductance is reduced and the surface flux noise remains the same. As a result, impurity spins in the bulk may dominate the flux noise in superconducting qubits in the Pippard regime, such as the ones using aluminum superconductors with large electron mean free path.
\end{abstract}

\maketitle

\section{Introduction\label{sec:intro}}

In the past few years there has been astounding progress on the design of superconducting circuits for large scale quantum computing \cite{Clarke2008}. Specific purpose circuits for quantum annealing containing over two thousand qubits have been developed and are currently being benchmarked against classical algorithms \cite{Denchev2016}. Universal circuits containing nine qubits with over one thousand quantum logic gates were successfully demonstrated \cite{Barends2016}. 

While designing these circuits require careful control of circuit parameters such as self and mutual inductances, Josephson critical currents, capacitances,  etc,  most of the design has so far been done without numerical prediction of circuit parameters. This has happened because the tools for numerical modeling of superconducting circuits are much more scarce \cite{InductEx2017}, in contrast to the tools available to the semiconductor industry. 

A key difference between superconductors and normal conductors is that in the former electrical current is carried by a condensate of Cooper pairs, a quantum superposition of many paired electrons and phonon modes that manifest macroscopic quantum behavior. As a consequence, the current-carrying entities (the charge of each Cooper pair) can not be localized within less than a length scale $\xi$, the uncertainty in the distance between the two electrons forming the Cooper pair, which is also known as Pippard coherence length. The impact of $\xi>0$ is that it makes the response of a superconductor to electromagnetic fields inherently nonlocal, in the sense that a field applied at a certain point $\bm{r}$ will affect the current density within a radius $\xi$ of $\bm{r}$. If we describe magnetic fields using a vector potential $\bm{A}(\bm{r},t)$ in the London gauge (defined by $\nabla \cdot \bm{A}=0$ with $\bm{A}\cdot\bm{\hat{n}} = 0$, where $\bm{\hat{n}}$ is the unit vector perpendicular to the surface of the superconductor), we get the following Pippard relation between current density and vector potential \cite{Pippard1953},
\begin{equation}
\bm{J}(\bm{r},t) = -\frac{3}{4\pi\xi_0\mu_{0}\lambda_{L}^{2}}\int_{{\rm SC}} \frac{\bm{R}[\bm{R}\cdot\bm{A}(\bm{r}',t)]}{R^4}
{\rm e}^{-\frac{R}{\xi}}
d^3 r',
\label{pippard_relation}
\end{equation}
where the integral is taken over the superconductor (SC), with $\bm{R}=\bm{r}-\bm{r}'$, and $\lambda_L$ the bare London penetration depth. Here $\xi_0$ is the Pippard coherence length for pure materials; $\xi$ is impacted by the presence of impurities according to $\frac{1}{\xi} = \frac{1}{\xi_0} + \frac{1}{l}$, where $l$ is the mean free path for electrons in the normal state ($l<\infty$  due to effects such as electron-impurity scattering).


A few remarks about the Pippard relation Eq.~(\ref{pippard_relation}) are in order: (1) It provides a good approximation to the more sophisticated methods using quantum field theory. For example, a formulation based on averaging out the Gor'kov equations over atomic length scales leads to a relation similar to Pippard's, except for the presence of an additional sum over Matsubara frequencies and an integral over electron velocities (See Eq.~(2.22) in \cite{Belzig1999}). 
(2) It describes AC fields provided that their frequency is lower than the superconducting gap, and is not resonant with impurity sub-gap states. This approximation corresponds to the dissipationless response in the so called ``two-fluid approximation'' \cite{Tinkham1996} (note that if we take the time derivative of Eq.~(\ref{pippard_relation}) and substitute $\bm{E}=-\partial_t \bm{A}$ we get the Pippard generalization of the first London relation). (3) The integral over $\bm{r}'$ in Eq.~(\ref{pippard_relation}) has a sharp cut off at the surface of the superconductor.  This prescription is equivalent to assuming diffusive scattering of electrons at the surface, i.e. the surface is assumed to be rough so that electrons coming from the surface have no memory of any previous exposure to the vector potential.

When $A(\bm{r},t)$ is approximately constant over the range $\xi$, the integrand can be approximated by a delta function and the Pippard relation reduces to the second London relation,
\begin{equation}
\bm{J}(\bm{r},t) = -\frac{1}{\mu_0\lambda^{2}}\bm{A}(\bm{r},t),
\label{london_relation}
\end{equation}
with $\lambda=\lambda_L\sqrt{\xi_0/\xi}$ the London penetration depth for impure superconductors. Since $\lambda$ is inversely proportional to the square root of the Cooper pair density, this important relation can be interpreted as a reduction of the effective Cooper pair density due to the lower coherence length $\xi$. Equation~(\ref{london_relation}) is ``local'' in the sense that the current density at any given point $\bm{r}$ within the superconductor responds only to the vector potential at that same point $\bm{r}$.

All modeling of superconducting devices to date has been done using the local London relation Eq.~(\ref{london_relation}) \cite{InductEx2017, Rhoderick1962, Lee1994, Anton2013b}. This can only be justified when $\xi$ is smaller than the characteristic length scale of $\bm{A}(\bm{r},t)$, which is set either by the smallest linear length scale of the wire, or by $\lambda$, whichever is smaller. 

Table~\ref{scparameters} shows the zero-temperature bare penetration depth $\lambda_L$ and Pippard coherence length $\xi_0$ for some common superconducting materials. 

\begin{table}[ht]
\centering
\caption{Zero-temperature ``bare" London penetration depth $\lambda_L$ and intrinsic Pippard coherence length $\xi_0$ for pure superconductors \cite{VanDuzer1999}.}\label{scparameters}
\begin{tabular}{c c c} 
\hline\hline
Superconductor & $\lambda_L$ $({\rm nm})$ & $\xi_0$ $({\rm nm})$ \\
 \hline
 Al & 16 & 1600  \\ 

 In & 19 & 490 \\

 Nb & 39 & 38 \\

 Pb & 37 & 83  \\

 Sn & 35 & 250  \\
\hline\hline
\end{tabular}
\end{table}

We see from Table~\ref{scparameters} that the Pippard coherence length is much larger than the penetration depth for most common superconductors. In the regime of large wire geometry with $\xi\gg \lambda$ this effect can be taken into account through a simple renormalization of the penetration depth, implying that local electrodynamics is still valid with a simple substitution of $\lambda_R$ for $\lambda$ in Eq.~(\ref{london_relation}) (See e.g. Section 3.11 of \cite{Tinkham1996}):
\begin{equation}
\lambda_R \approx (\lambda^{2}\xi)^\frac{1}{3}.
\label{lambda_R}
\end{equation}

Here we perform exact numerical calculations of current density and vector potential for superconducting wires in the nonlocal regime. We show that local electrodynamics generally breaks down, impacting circuit parameters such as inductance and flux noise. 

The article is organized as follows. Section~\ref{Local} describes analytical solutions for the current density and vector potential inside a cylindrical wire using local electrodynamics. Section~\ref{nonlocal} presents our 
numerical method for exact solution of the self-consistent relations arising in nonlocal electrodynamics; 
Section~\ref{numerical_results} describes our explicit numerical results, their comparison to the local case, and to simple approximations based on $\lambda_R$. Section~\ref{inductance} applies these results to calculations of internal and kinetic inductance, and Section~\ref{flux_noise} to calculation of flux noise due to impurities at the surface and the bulk of the wires. Finally, Section~\ref{conclusions} presents our conclusions. 

\section{\label{Local}Local electrodynamics: Exact analytical solution}

\begin{figure}[ht]
    \includegraphics[width = 0.49\textwidth]{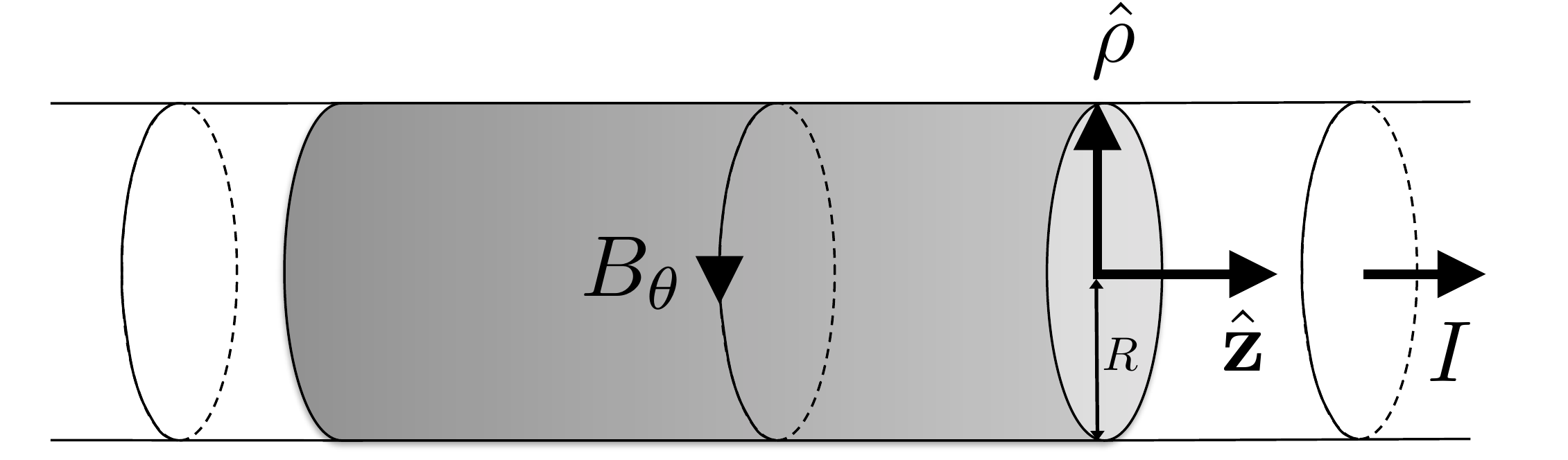}
    \caption{Infinite wire with cylindrical geometry. The current is assumed to flow along $+\hat{\bm{z}}$.}
    \label{FigCylinder}
\end{figure}

A well known challenge in the computation of superconducting current density and the Meissner effect is the presence of singularities at the wire's edge. Even for the simpler local case, these singularities require the use of uncontrolled approximations, such as the ones made for thin films and strip lines \cite{Cooper1961, Marcus1961, Rhoderick1962, Lee1994, VanDuzer1999, Laforest2015, Anton2013b}. 
Here we avoid edge singularities by focusing on a much simpler wire geometry, the infinitely long, straight, and cylindrical superconducting wire shown in Fig.~\ref{FigCylinder}.  As we show here, the choice of an edgeless geometry greatly simplifies the calculations and makes the nonlocal case exactly solvable, without the use of approximations that are hard to justify. 
The cylindrical wire is not just an idealization: It provides a realistic model for the coaxial cable with return current flowing on an external cylinder concentric with the wire (the return current produces zero magnetic field inside the wire, and does not affect its SC current density). 

Combining the Maxwell equation 
\begin{equation}
\left(\nabla^2-\frac{1}{c^2}\partial^{2}_{t}\right)\bm{A}=-\mu_0 \bm{J}
\label{maxwell}
\end{equation}
with Eq.~(\ref{london_relation}) we get the usual equations describing local electrodynamics,
\begin{equation}
\left(\nabla^2-\frac{1}{c^2}\partial^{2}_{t}\right)
\left(
\begin{array}{c}
\bm{A}\\
\bm{J}
\end{array}\right)
=\frac{1}{\lambda^{2}}
\left(
\begin{array}{c}
\bm{A}\\
\bm{J}
\end{array}\right).
\label{local_eqns}
\end{equation}
Assuming an AC current density parallel to the $z$-axis, we get $\bm{J}(\bm{r},t) = J(\rho)\textrm{e}^{-i\omega t}\hat{\bm{z}}$, because $J$ can not depend on the polar angle $\theta$ due to the cylindrical symmetry. In the London gauge the vector potential assumes a similar form, $\bm{A}(\bm{r},t) = A(\rho)\textrm{e}^{-i\omega t}\hat{\bm{z}}$. Plugging this into Eq.~(\ref{local_eqns}) and noting that $\lambda \ll c/\omega$ in all cases of interest ($\omega/2\pi<100$~GHz) leads to the following exact solutions in terms of the modified Bessel functions of the first kind $I_n(x)$:
\begin{equation} \label{eq:local_A}
A_{{\rm local}}(\rho) = -\frac{\lambda \mu_0 I}{2\pi R}\frac{I_0(\rho/\lambda)}{I_1(R/\lambda)}, \quad 0\leq\rho\leq R,
\end{equation}
and 
\begin{equation} \label{eq:local_J}
J_{{\rm local}}(\rho) = \frac{I}{2\pi R\lambda}\frac{I_0(\rho/\lambda)}{I_1(R/\lambda)}, \quad 0\leq\rho\leq R.
\end{equation}
Here $R$ is the radius of the wire, and the expressions are normalized by the total current $I$.
The internal magnetic field $\bm{B}_{{\rm local}}=B_{{\rm local}}(\rho)\textrm{e}^{-i\omega t}\hat{\bm{\theta}}$ is found by taking the curl of vector potential, or just using Amp\`{e}re's law,
\begin{equation} \label{eq:local_B}
	B_{{\rm local}}(\rho) = \frac{\mu_0 I}{2\pi R}\frac{I_1(\rho/\lambda)}{I_1(R/\lambda)}, \quad 0\leq\rho\leq R. 
\end{equation}
These equations will be the baseline for comparison with the nonlocal current densities and vector potentials calculated in the next section.

\section{Exact numerical method to treat nonlocal electrodynamics\label{nonlocal}}

In the case where the spatial variation of the field is smaller than the effective size of the Cooper pairs (defined by the parameter $\xi$), nonlocal effects must be taken into account. 
Since all the fields are independent of $z$ (the wire extends infinitely in the $\hat{\bm{z}}$ direction), we can use an effective two-dimensional Pippard relation,
\begin{eqnarray}
J(\rho) &=&  -\frac{1}{2\pi\xi\mu_{0}\lambda^{2}}\int A(\rho')\frac{\exp(- \frac{\mid \bm{r}_{\bot}-\bm{r}'_{\bot} \mid}{\xi})}{\mid \bm{r}_{\bot}-\bm{r}'_{\bot} \mid}d^2 r' , \nonumber\\
 &=& -\frac{1}{\mu_0\lambda^{2}} \int_{0}^{R} d\rho'\rho' K_{P}(\rho,\rho')A(\rho').\label{2DPipp}
\end{eqnarray}
The relation is conveniently written as an integral over the Pippard Kernel defined by 
\begin{equation}
K_P(\rho,\rho')=\frac{1}{2\pi \xi}\int_{0}^{2\pi} \frac{\exp(- \frac{\mid \bm{r}_{\bot}-\bm{r}'_{\bot} \mid}{\xi})}{\mid \bm{r}_{\bot}-\bm{r}'_{\bot} \mid}d\theta'.
\end{equation}
Note that $K_P$ is independent of $\theta$ because $|\bm{r}_{\bot}-\bm{r}'_{\bot}|$ is a function of $(\theta-\theta')$, so the integration over $\theta'$ fully erases the $\theta$ dependence. The Kernel becomes a delta function when $\xi\rightarrow 0$,
\begin{equation}
K_{P}(\rho,\rho') \underset{\xi \to \, 0}{=} \frac{\delta(\rho-\rho')}{\rho'}. 
\end{equation}

To solve for the vector potential and current density in the nonlocal scenario, we formulate the problem as a Poisson equation with Dirichlet boundary conditions,
\begin{equation}
\nabla^2 A(\rho) = -\mu_0 J(\rho)\;,\; A(\rho=R) = A_0.
\label{Poisson_A}
\end{equation}
Here $A_0$ is the value of the vector potential at the surface of the wire. This choice of boundary condition turns out to be quite convenient for numerical calculations. 
Later we will show how to convert this solution into the desired boundary condition of constant $B=-\partial_\rho A$ at the surface (corresponding to a value of total current $I$). 

\begin{figure}[t]
    \includegraphics[width = 0.49\textwidth]{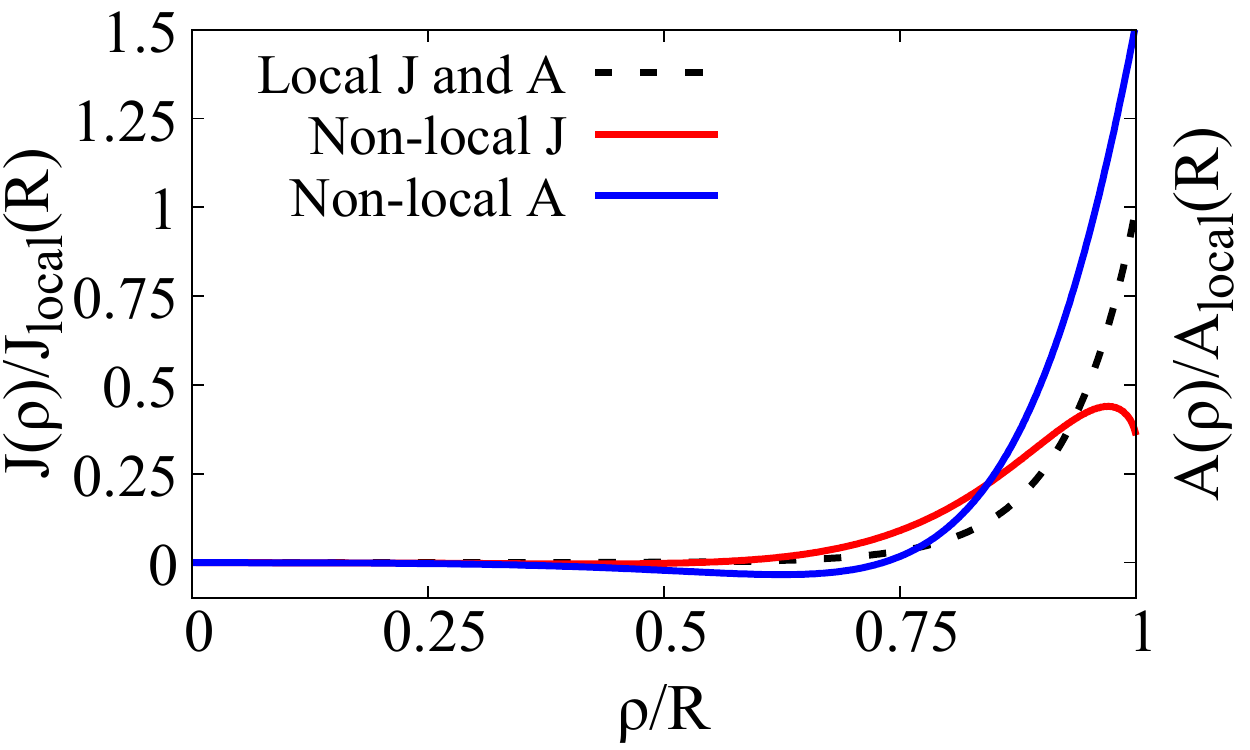}
    \caption{Comparison between local and nonlocal current density and vector potential for $R=1000$~nm, $\lambda=70$~nm, and $\xi=200$~nm. Note that $J$ and $A$ are normalized by their local values so that in the local regime they fall on the same curve. The impact of nonlocality is evident in two features: $J$ acquires a peak away from the surface of the wire, and it changes sign inside the wire. Both $A$ and the magnetic field $B_\theta=-\partial_\rho A$ also change sign inside the wire, due to self-induced overscreening.\label{local_vs_nonlocal}}
\end{figure}

An explicit solution to Eq.~(\ref{Poisson_A}) can be obtained using the Dirichlet Green's function for the 2D Laplace operator; this Green's function satisfies
\begin{subequations}
\begin{eqnarray}
\nabla^{2} G(\rho,\theta,\rho',\theta') &=& \frac{\delta(\rho-\rho')\delta(\theta-\theta')}{\rho'},\\
G(\rho,\theta,\rho'=R,\theta')&=&0.
\label{DirichletG}
\end{eqnarray}
\end{subequations}
The fact that $G$ is exactly equal to zero at the surface of the wire leads to the following Green's identity,
\begin{eqnarray}
A(\rho)&=&-\mu_0 \int_{0}^{R}d\rho' \rho' \int_{0}^{2\pi}d\theta' G(\rho,\theta,\rho',\theta')J(\rho')\nonumber\\
&&+R \int_{0}^{2\pi}d\theta' (\partial_{\rho'}G) A(\rho')\Big|_{\rho' =R},
\label{GreensIdentity}
\end{eqnarray}
where the second term in the RHS is a surface term evaluated at $\rho'=R$. 

We now show two results that greatly simplify our method. First, there is an explicit analytical expression for the Green's function satisfying Eq.~(\ref{DirichletG}) (See Example 13.3 in \cite{Trim1990}):
\begin{eqnarray}
&&G(\rho,\rho', \theta-\theta')\nonumber\\ 
&&=\frac{1}{4\pi}\ln{\left[\frac{\rho^2+\rho'^2 -2\rho\rho'\cos(\theta-\theta')}{\rho^2\rho'^2/R^2+R^2-2\rho\rho'\cos(\theta-\theta')}\right]}.
\label{Gexp}
\end{eqnarray}
The second simplifying result is that the surface term in Eq.~(\ref{GreensIdentity}) can be shown to be exactly equal to $A_0$, the value of the vector potential at the surface:
\begin{eqnarray}
&&R \int_{0}^{2\pi}d\theta' (\partial_{\rho'} G) A(\rho')\Big|_{\rho' =R}\nonumber\\
&&=A_0\int_{0}^{2\pi} \frac{1}{2\pi}\left[\frac{1-\rho^2}{\rho^2+1-2\rho\cos(\theta-\theta')}\right]d\theta' \nonumber\\
&&= A_0.
\end{eqnarray}
The last identity can be checked with {\it Mathematica}.
Altogether the equation for $A(\rho)$ becomes,
\begin{equation}
A(\rho) = A_0 -\mu_0\int_{0}^{R}d\rho' \rho'\int_{0}^{2\pi}d\theta' G(\rho,\rho', \theta')J(\rho').
\end{equation}

\begin{figure*}[t]
\begin{center}
\subfloat[]{\includegraphics[width=0.49\textwidth]{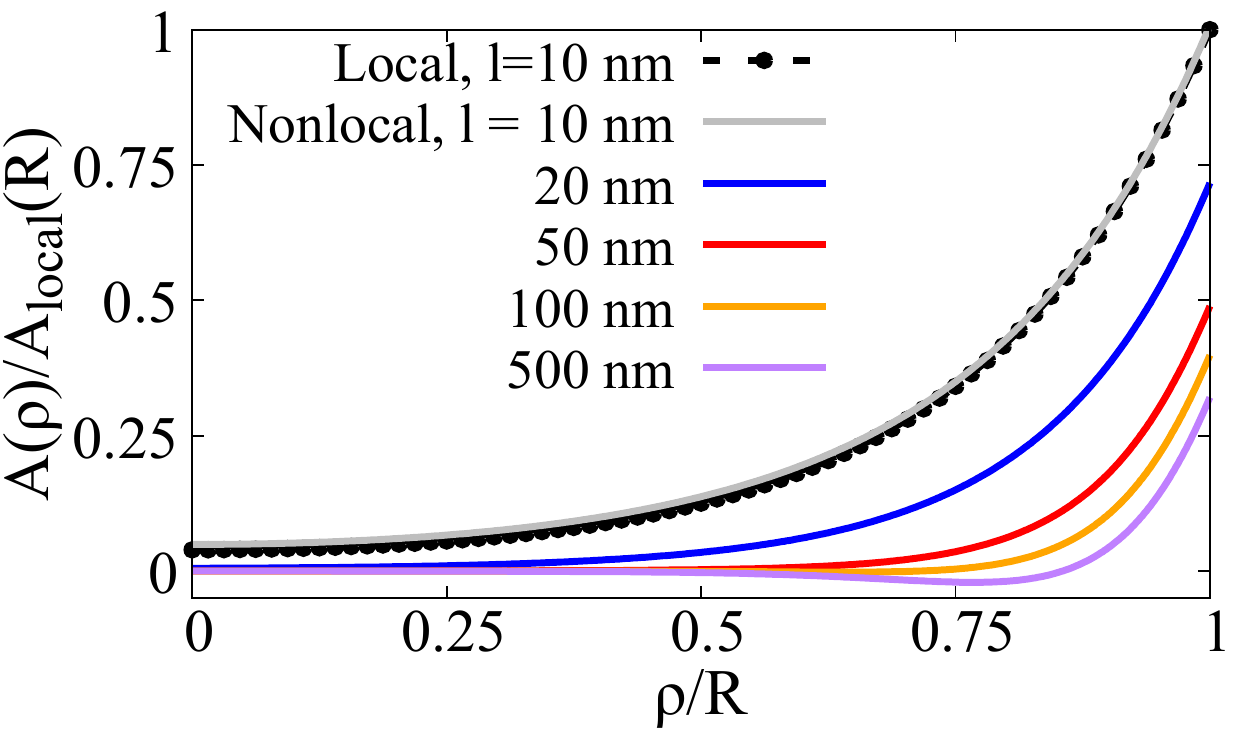}
\label{Amultixi}}
\subfloat[]{\includegraphics[width=0.49\textwidth]{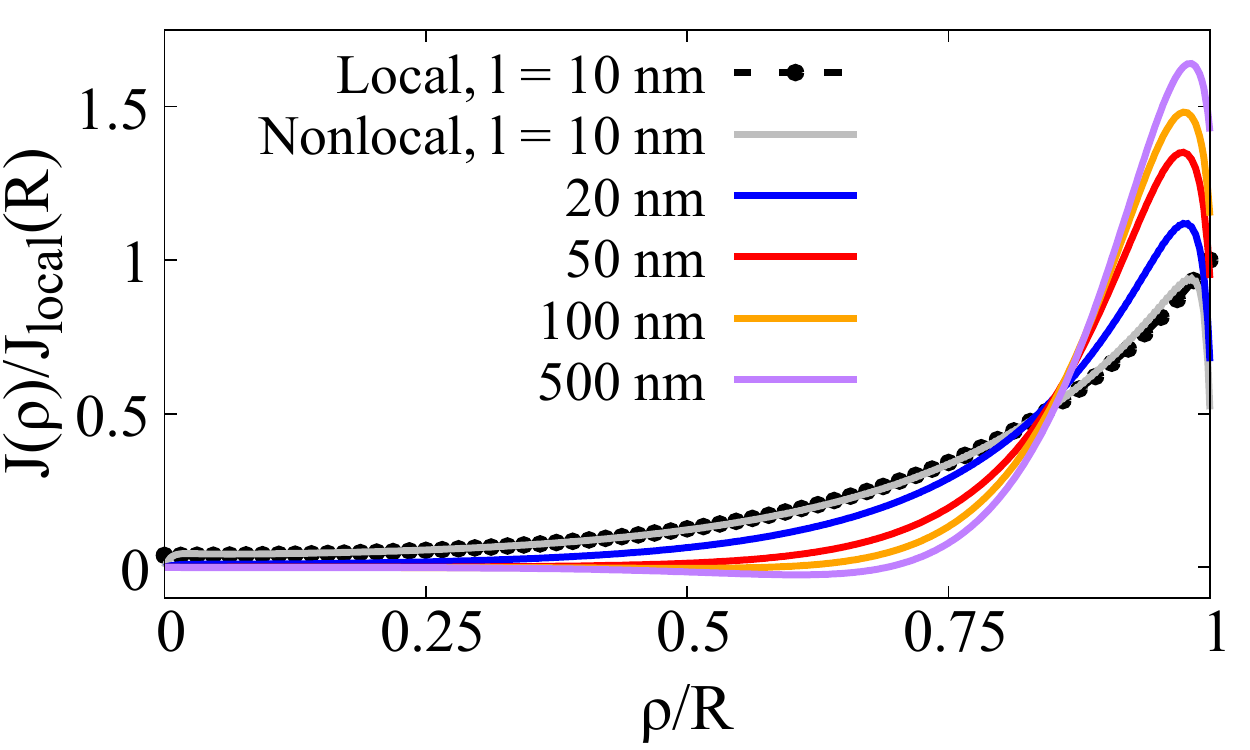}
\label{Jmultixi}}
\end{center}
\caption{(Color online) (a) Vector potential and (b) current density for $R = 1000$~nm and different values of mean free path $l$. We used parameters appropriate for aluminum, $\lambda_L=16$~nm and $\xi_0=1600$~nm, with $\xi=\xi_0 l/(\xi_0 + l)$ and $\lambda=\lambda_L\sqrt{\xi_0/\xi}$ changing with $l$. 
The plots are normalized by the local surface values $A_{{\rm local}}(R)$ and $J_{{\rm local}}(R)$ calculated from Eqs.~(\ref{eq:local_A})~and~(\ref{eq:local_J}) with parameter $\lambda=203$~nm fixed for all curves (this is the same $\lambda$ used in the nonlocal $l=10$~nm case).
Note that the surface magnetic field $B_{\theta}(R)$ and the total current $I$ is the same for all curves. A plot of the magnetic field $B_\theta(\rho)=-\partial_\rho A$ is shown in Fig.~\ref{Fig_FluxVectorR1000nm}.\label{multixi}}
\end{figure*}

We can now replace $J(\rho')$ with the 2D Pippard relation defined in Eq.~(\ref{2DPipp}):
\begin{equation} \label{Fredholm}
A(\rho) = A_{0}  + \int_{0}^{R}d\rho'\rho'\int_{0}^{R}d\rho'' \rho'' K(\rho,\rho',\rho'')A(\rho''),
\end{equation}
where the kernel $K(\rho,\rho',\rho'')$  intertwines the Green's function with the Pippard relation,
\begin{equation}
K(\rho,\rho',\rho'') = \frac{1}{\lambda^{2}}\left[\int_{0}^{2\pi}d\theta' G(\rho,\rho', \theta')\right] K_P(\rho',\rho'').
\end{equation}

Equation~(\ref{Fredholm}) is a Fredholm integral equation, solvable through a variety of numerical methods.  
The numerical solutions presented in the following section were obtained with FIE, the Fredholm integral equation solver developed by Atkinson et al. \cite{Atkinson2007}. 
It uses the Nystr\"{o}m method, which discretizes  into a mesh $\{\rho_i\}$ so that the integral part can be converted into a matrix times the vector $\{A(\rho_i)\}$. Equation~(\ref{Fredholm}) then becomes 
an inhomogeneous linear system of equations, with the set of $A(\rho_i)$'s as the unknowns, and inhomogeneity equal to $A_0$. The system can be solved numerically with matrix inversion, and each $A(\rho_i)$ becomes proportional to $A_0$. 

Once $A(\rho)$ is obtained, $J(\rho)$ can be calculated by numerical integration of the Pippard relation Eq.~(\ref{2DPipp}), and the total current $I$ becomes a constant times $A_0$. 
Hence, $A_0$ is the constant of integration that sets the total current $I$ flowing through the wire. 

This observation leads to a simple method to obtain the vector potential and current density with boundary condition set by the magnetic field at the surface, $B_0=\mu_0 I/(2\pi R)$. To do this, we set $A_0=1$ and the output of the Fredholm solver yields the function $A'(\rho)=A(\rho)/A(R)$. We do not know the value of $A(R)$ a priori, so we compute $\partial_\rho A'(R)$ numerically and consider the renormalized function:
\begin{eqnarray}
A''(\rho)&=& \frac{I_1(R/\lambda)}{\lambda I_0(R/\lambda)}\frac{A'(\rho)}{\partial_\rho A'(R)}=\frac{-B_0}{A_{{\rm local}}(R)}\frac{A(\rho)}{\partial_\rho A(R)}\nonumber\\
&=& \frac{A(\rho)}{A_{{\rm local}}(R)}.
\label{eq:App}
\end{eqnarray}
In the last identity we used the fact that $B_0=-\partial_\rho A(R)$ is the same for the local and nonlocal cases (Amp\`{e}re's law). The function $A''(\rho)$ is our desired result: It is the nonlocal vector potential in a wire with total current $I$ in units of a known function, $A_{{\rm local}}(R)$ [Eq.~(\ref{eq:local_A})]. The associated current density is given by 
\begin{equation}
J''(\rho)=\frac{J(\rho)}{J_{{\rm local}}(R)}=\int_{0}^{R}d\rho' \rho' K_P(\rho,\rho')A''(\rho').
\label{eq:Jpp}
\end{equation}

\section{Vector potential and current density in the nonlocal regime\label{numerical_results}}

We now show exact numerical calculations of vector potential and current density in the cylindrical wire, using the method described in Sec.~\ref{nonlocal}. 
Section~\ref{comp_renormalized} compares our exact solutions to a much simpler numerical approximation based on the renormalized penetration depth of Eq.~(\ref{lambda_R}).

\subsection{Exact numerical results}

Figure~\ref{local_vs_nonlocal} compares local and nonlocal calculations of $A(\rho)$ and $J(\rho)$ for $R=1000$~nm, $\lambda=70$~nm, and $\xi=200$~nm. Two features of the nonlocal regime are evident: 
$J$ acquires a peak away from the surface of the wire and changes sign inside the wire. As a result, $J$ is no longer directly proportional to $A$, signaling the break down of the London relation Eq.~(\ref{london_relation}). 
Both $A$ and the magnetic field $B_\theta=-\partial_\rho A$ change sign inside the wire. This phenomena is called overscreening and is the ``smoking gun'' for detecting nonlocality \cite{Pippard1953}. It occurs because the screening current of the superconductor is rigid on the scale of $\xi$, so it ends up producing a magnetization that is larger than the internal magnetic field, leading to overscreening of the field. Overscreening has been demonstrated experimentally in cylindrical SC films subject to an external magnetic field \cite{Drangeid1962}. In contrast, Fig.~\ref{local_vs_nonlocal} shows \emph{self-induced overscreening}, in that it occurs 
solely due to the fields produced by the supercurrent, without an applied external field.

Figure~\ref{multixi} shows calculations of $A$ and $J$ for $R=1000$~nm and different values of the electron mean free path $l$. We used parameters for aluminum: $\lambda_L=16$~nm, $\xi_0=1600$~nm, 
with $\xi=\xi_0 l/(\xi_0 + l)$ and $\lambda(l)=\lambda_L\sqrt{\xi_0/\xi}$ changing with $l$. 
It should be noted that while each curve has a different $\lambda(l)$, all curves are normalized by the local surface values $A_{{\rm local}}(R)$, $J_{{\rm local}}(R)$ with $\lambda$ fixed at $\lambda(10~{\rm nm})=203$~nm. To do this, one must choose the normalization factor in Eq.~(\ref{eq:App}) to be 
$I_1(R/\lambda(10~{\rm nm}))/[\lambda(10~{\rm nm}) I_0(R/\lambda(10~{\rm nm}))]$, and also multiply Eq.~(\ref{eq:Jpp}) by $[\lambda(10~{\rm nm})/\lambda(l)]^2$. Doing this ensures that all curves are plotted with the same surface magnetic field $B_0$ and total current $I$.

When $l \rightarrow 0$, both $A$ and $J$ converge to the local solution, again plotted with $\lambda(10~{\rm nm})$.  The local $J$ is always peaked at the edge of the wire,  
decreasing exponentially towards the interior. In contrast, nonlocal $J$ is peaked off edge, with non-exponential fall off towards the interior.  
Remarkably, the off-edge peak appears even for small values of $\xi \approx l$ ($l=10$~nm in Fig.~\ref{Jmultixi}), showing that even a small degree of nonlocality qualitatively impacts current density.
All superconductors have some degree of nonlocality, so that a significant portion of the current flows away from the surface.

\begin{figure}[ht]
    \includegraphics[width = 0.49\textwidth]{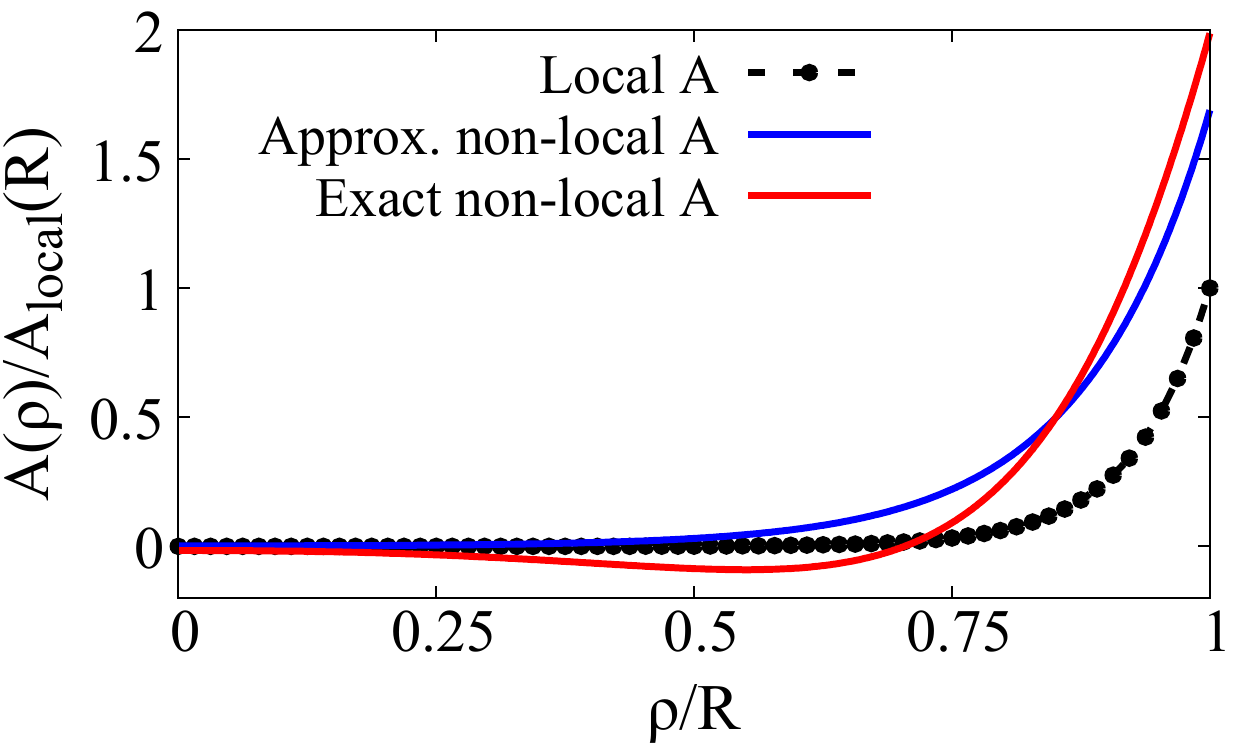}
    \caption[]{Compares the vector potential calculated using the local approximation with $\lambda_R=0.85 (\lambda^{2}\xi)^{1/3}$ to the exact solution of the Fredholm integral equation, for  $\lambda=70$~nm and $\xi = 500$~nm.\label{figAmod}}
\end{figure}

\begin{figure}[ht]
    \includegraphics[width = 0.49\textwidth]{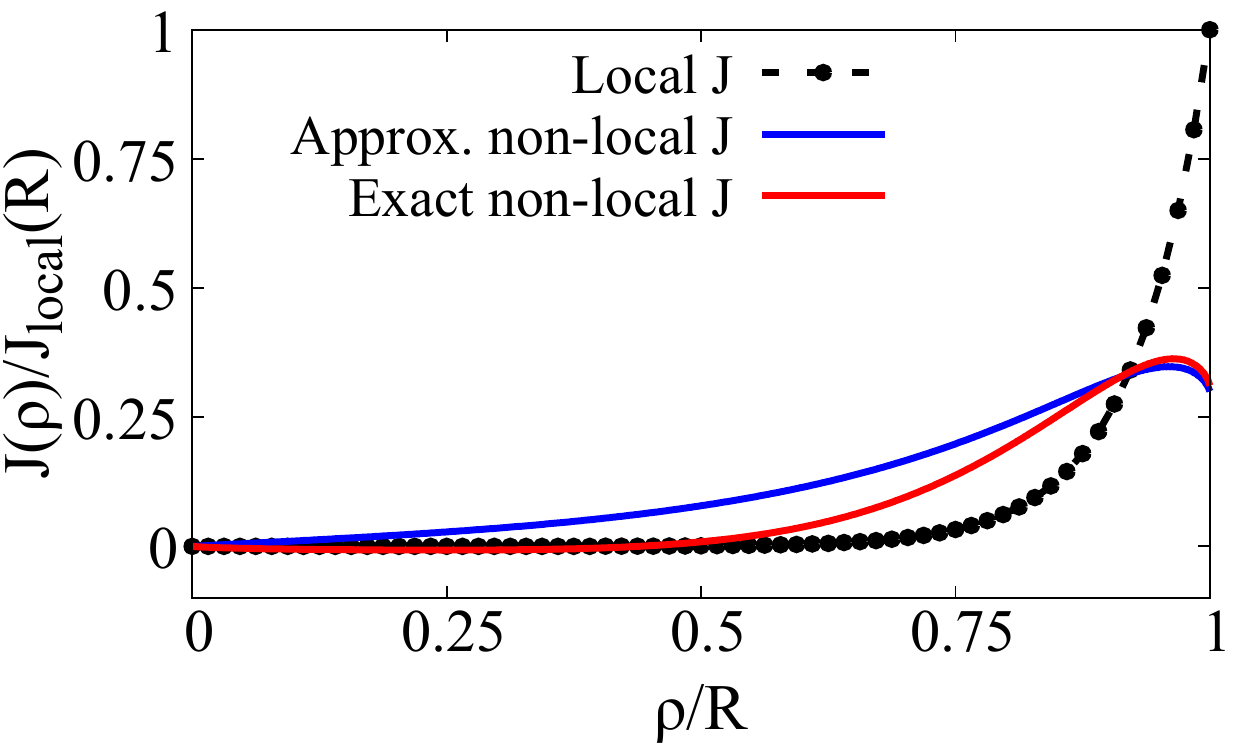}
    \caption[]{Approximate current density obtained by numerical evaluation of the Pippard relation Eq.~(\ref{pippard_relation}) using the approximate $A(\rho)$ of Fig.~\ref{figAmod}.\label{figJmod}}
\end{figure}

\subsection{Comparison to local approximation with renormalized penetration depth \label{comp_renormalized}}

As noted in Section~\ref{sec:intro}, in the limit of a large superconductor the Pippard relation can be approximated by a London relation with renormalized penetration depth.  Figure~\ref{figAmod} shows the vector potential obtained from the solution of Eq.~(\ref{london_relation}) using
$\lambda_R= 0.85\times (\lambda^{2}\xi)^{1/3}$ in place of $\lambda$. Such a choice of $\lambda_R$ is seen to yield a good description of $A$ in the cylindrical geometry for a wide range of $\lambda, \xi$.
We take this approximate vector potential and plug into the Pippard relation Eq.~(\ref{2DPipp}) to obtain an approximate current density. Figure~\ref{figJmod} compares this approximate $J$ to the exact solution. 

The local approximation with $\lambda_R$ does not describe overscreening, and overestimates the value of the total current, in spite of having the same magnetic field of the exact solution at $\rho=R$. This occurs because the approximate solution does not satisfy the Maxwell equation together with the Pippard relation. Apart from these shortcomings it provides a reasonable description of nonlocal electrodynamics for large $\xi$, especially near the edge of the superconducting wire. 

\section{Inductance in the nonlocal regime\label{inductance}}

We now apply our theory to nonlocal calculations of kinetic and internal (magnetic) inductance. These are important device quantities in that their sum determines the response time or ``inertia'' of the circuit, with kinetic inductance playing a key role in devices for single photon detection \cite{Mazin2002}. Kinetic inductance depends crucially on the wire's current distribution \cite{VanDuzer1999},
\begin{equation}
L_k = \frac{\mu_0\lambda^{2}}{I^2}\int_{{\rm SC}}\bm{J}^2(\bm{r}) d^3 r.
\label{Lk}
\end{equation}
In contrast, internal inductance depends on the magnetic field inside the wire, 
\begin{equation}
L_{{\rm int}} = \frac{1}{\mu_0 I^2}\int_{{\rm SC}}\bm{B}^2(\bm{r}) d^3 r.
\end{equation}

In the local case the inductances can be calculated by plugging Eq.~(\ref{eq:local_J}) for the current density and Eq.~(\ref{eq:local_B}) for the magnetic field. After performing the integrals the local inductances for a wire of length $l$ become
\begin{subequations}
\begin{eqnarray}
L_{k}^{{\rm local}} &=& \frac{\mu_0 l}{2\pi R^2 I_1^2(R/\lambda)} \int_0^R d\rho \rho I_{0}^{2}(\rho/\lambda), \\
L_{{\rm int}}^{{\rm local}} &=& \frac{\mu_0 l}{2\pi R^2 I_{1}^{2}(R/\lambda)} \int_{0}^{R}d\rho \rho I_1^2(\rho/\lambda).
\end{eqnarray}
\end{subequations}

\begin{figure}[ht]
    \includegraphics[width = 0.49\textwidth]{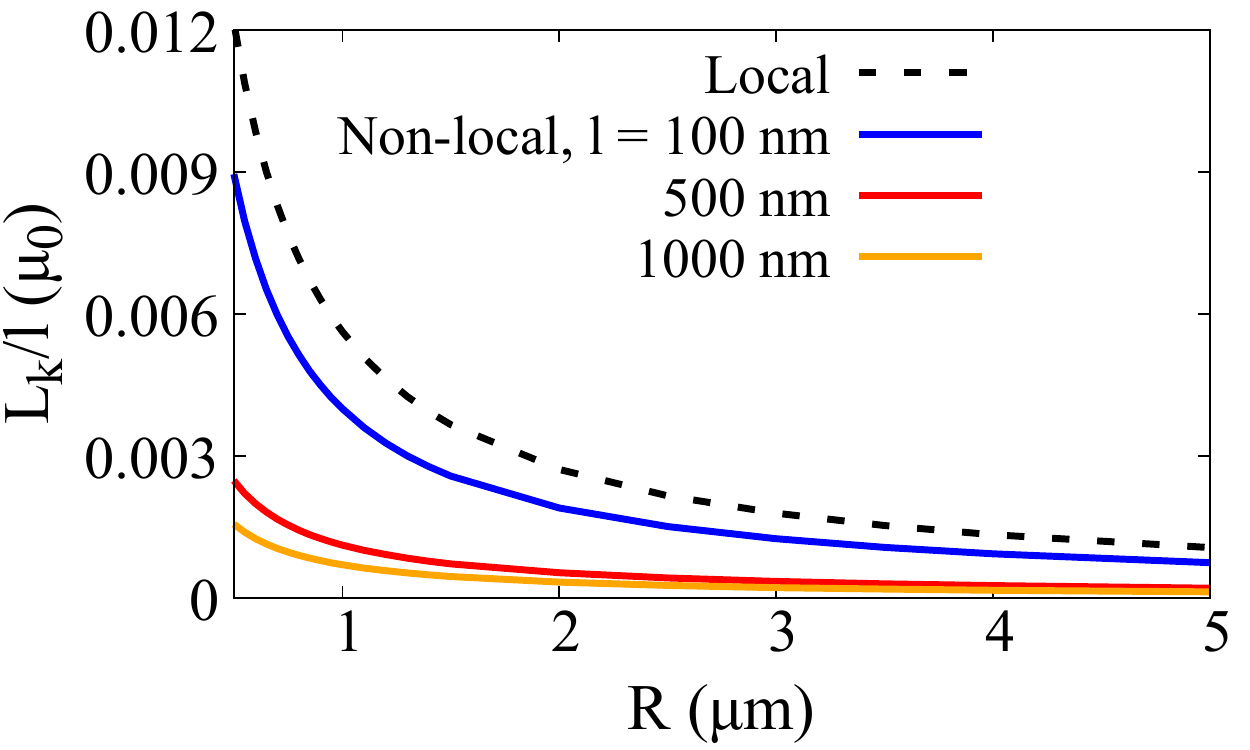}
    \caption{Kinetic inductance per unit length as a function of wire radius, for the local case (dashed), and nonlocal cases with varying $l$. The parameters $\lambda, \xi$ are the same used in Fig.~\ref{multixi}.\label{figLk}}
\end{figure}

\begin{figure}[ht]
    \includegraphics[width = 0.49\textwidth]{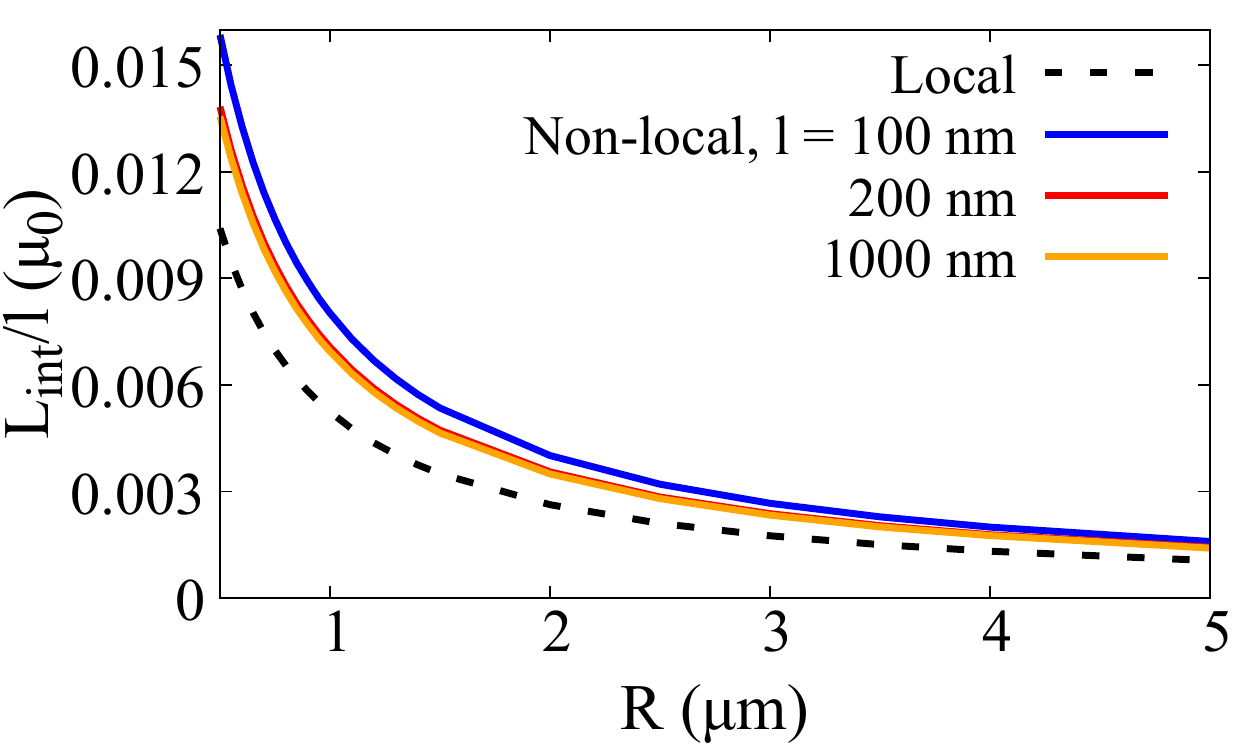}
    \caption[$L_{int}$ nonlocal Infinite Cylinder: 2D Pippard]{Internal inductance per unit length as a function of wire radius, for the local case (dashed), and nonlocal cases with varying $l$. The parameters $\lambda, \xi$ are the same used in Fig.~\ref{multixi}.\label{figLint}}
\end{figure}

Figures~\ref{figLk}~and~\ref{figLint} compare these local results to explicit numerical calculations of the inductances in the nonlocal regime, with the same $\lambda$ and $\xi$ used in Fig.~\ref{multixi}.
From these plots it is clear that kinetic inductance decreases with increasing mean free path $l$; in contrast, internal inductance is non-monotonic with increasing $l$.

\begin{figure}[ht]
    \includegraphics[width = 0.49\textwidth]{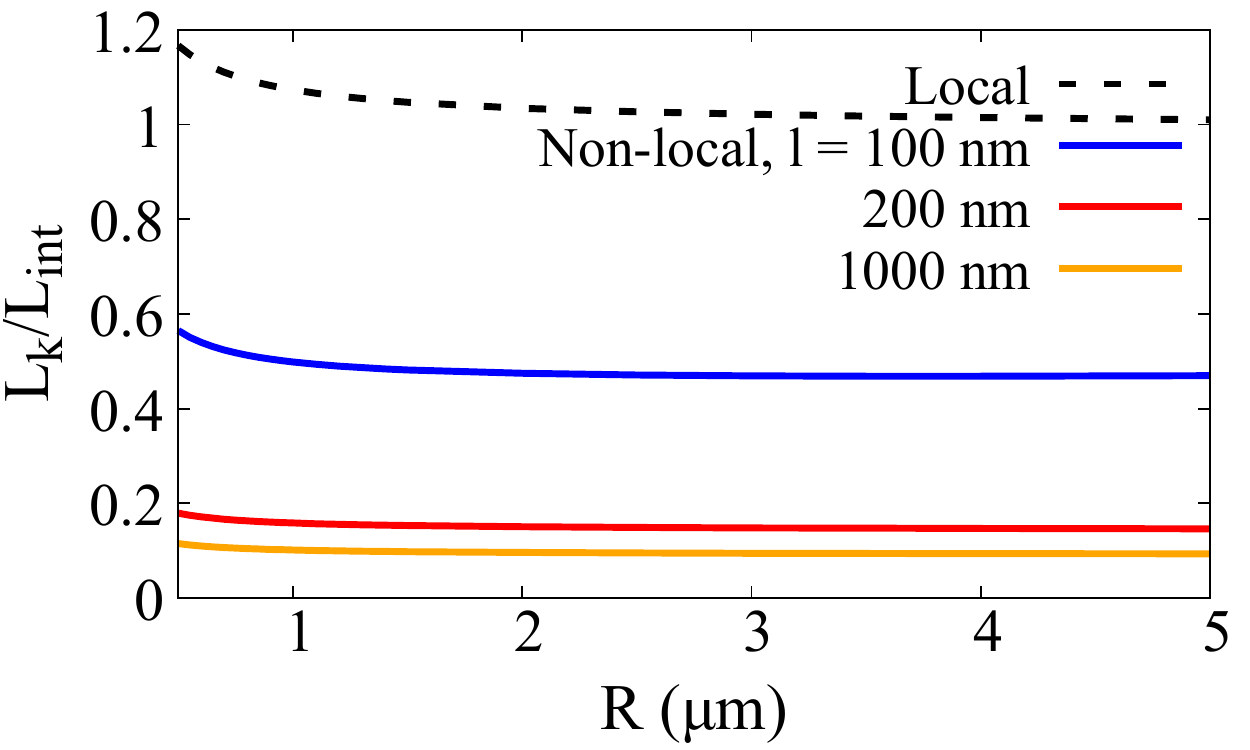}
    \caption[$L_k/L_{int}$ nonlocal Infinite Cylinder: 2D Pippard]{Ratio of kinetic and internal inductances as a function of wire radius, for the local case (dashed), and nonlocal cases with varying $l$. The parameters $\lambda, \xi$ are the same used in Fig.~\ref{multixi}. In the local regime kinetic inductance is always larger than internal inductance; in contrast, the nonlocal regime shows the opposite behaviour.\label{RatioInductances}}
\end{figure}

In the local regime kinetic inductance is always larger than internal inductance. However as the wire radius increases the inductances become approximately equal to each other. This result is similar to calculations performed on superconducting thin films \cite{VanDuzer1999}.  In contrast, the nonlocal regime shows the opposite behaviour: Internal inductance is always larger than kinetic inductance, and this disparity increases as $l$ or $\xi$ increases. The ratio of inductances is plotted in Fig.~\ref{RatioInductances}.

\begin{figure}[ht]
    \includegraphics[width = 0.49\textwidth]{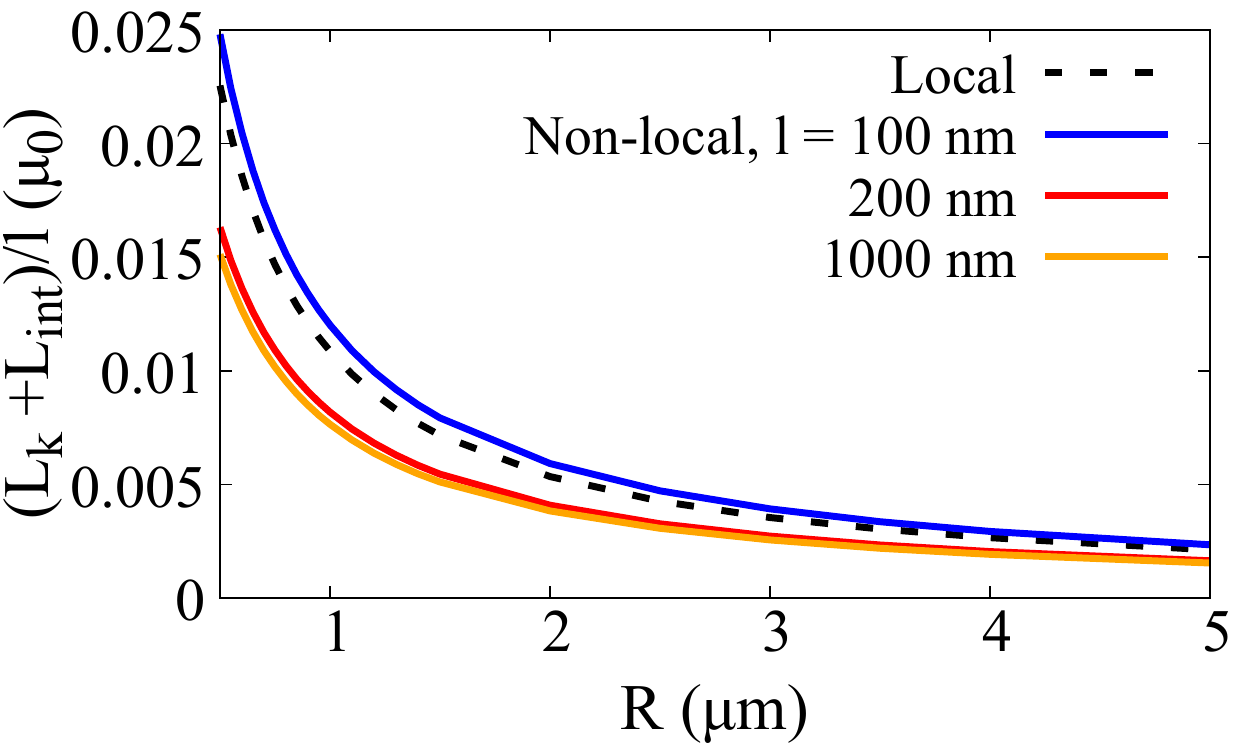}
    \caption[Sum of kinetic and internal inductances $L_k+L_{int}$ for nonlocal Infinite Cylinder: 2D Pippard]{Total inductance as the sum of kinetic and internal inductances, as a function of wire radius, for the local (dashed) and nonlocal cases with varying $l$. The parameters $\lambda, \xi$ are the same used in Fig.~\ref{multixi}.\label{SumInductances}}
\end{figure}

Finally, Fig.~\ref{SumInductances} shows the sum of kinetic and internal inductances; as $l$ increases from zero the total nonlocal inductance becomes larger than the local one; increasing $l$ further makes nonlocal inductance smaller than the local one.

\section{Flux Noise in the nonlocal regime\label{flux_noise}}

The performance of Superconducting Quantum Interference Devices (SQUIDs) and other superconducting circuits is limited by the presence of intrinsic flux noise, whose origin is believed to be due to the time dependent fluctuation of spin impurities located within the surfaces and interfaces forming the device \cite{Koch1983, Wellstood1987, Koch2007, deSousa2007, Sendelbach2008, Lanting2014, Kumar2016}. 
Consider a set of localized impurities labeled by $i=1,2, \ldots, N$. Each impurity is located at position $\bm{R}_i$, and its magnetic moment is described by the dimensionless spin operator $\bm{s}_i$. 
The flux sensed by a circuit due to the presence of impurities can be written as \cite{Laforest2015}
\begin{equation}
\Phi=-\sum_i \bm{F}_i \cdot \bm{s}_i, 
\label{phi_f_s}
\end{equation}
with a flux vector $\bm{F}_i$ pointing along the magnetic field produced by the device's current density,
\begin{equation}
\bm{F}_i=\frac{g\mu_B}{I}\bm{B}(\bm{R}_i). 
\label{fi}
\end{equation}
Here $g\mu_B$ is the magnetic moment of the electronic impurity, with $\mu_B$ the Bohr magneton. 

\begin{figure}[ht]
   \includegraphics[width = 0.49\textwidth]{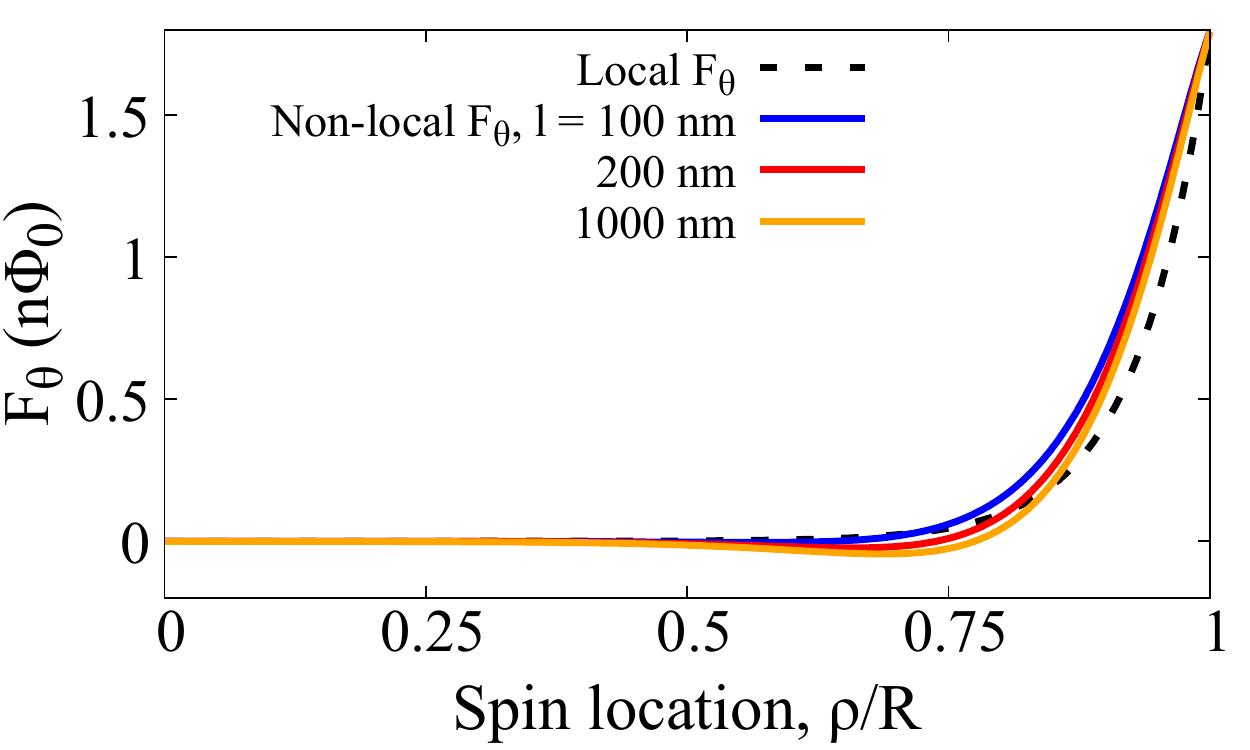}
   \caption[$F_\theta(\rho)$ Infinite Cylinder:  $R = 1000$ nm, 2D Pippard]{Flux vector as a function of impurity spin radial location $\rho$ inside the wire, for the same parameters used in Fig.~\ref{multixi}.  The flux vector determines the flux coupled to the wire according to $\Phi=-\bm{F}_i\cdot \bm{s}_i$, where $\bm{s}_i$ is the impurity spin operator.\label{Fig_FluxVectorR1000nm}}
\end{figure}

For a cylindrical wire we have $\bm{F}_i=F_{\theta}(\rho_i)\bm{\hat{\theta}}$, i.e. the flux vector only depends on the spin's radial position $\rho_i$. 
Figure~\ref{Fig_FluxVectorR1000nm} shows $F_{\theta}$ as a function of $\rho_i$ for the same parameters used in Fig.~\ref{multixi}.  Note that 
$F_{\theta}$ is always peaked at the surface, so the device is mostly sensitive to impurities located at the surfaces and interfaces. However, as the nonlocal case sets in the device becomes more sensitive to impurities in the bulk. 

We now turn to calculations of the total flux noise power (integrated over all frequencies) produced by the impurity spins, $\langle\left(\delta\Phi\right)^{2}\rangle=\langle \Phi^2\rangle -\langle\Phi\rangle^2$.
At large temperatures (larger than any ordering temperature for the spins) the flux noise power due to spins at the surface of the wire is given by \cite{Laforest2015}
\begin{eqnarray}
\langle \left(\delta\Phi\right)^{2}\rangle &=& \frac{S(S+1)}{3} \sigma_2 \int_{{\rm surface}} d^2 r \left| \bm{F}(\bm{r})\right|^{2},\nonumber\\
&=& \frac{S(S+1)(g\mu_B \mu_0)^{2}}{6\pi}\frac{\sigma_2 l}{R}.
\label{surfacefluxnoise}
\end{eqnarray}
Here $S$ is the spin quantum number of the impurity species, and $\sigma_2$ is the impurity areal density. In the second line we used Amp\`{e}re's law to compute the magnetic field at the surface of the wire. 
This exact result shows that the \emph{surface flux noise is the same for the local and nonlocal regimes}. Note how Eq.~(\ref{surfacefluxnoise}) scales proportional to $\sigma_2 l/R$, a result that is quite similar to 
the $\sigma_2 l/W$ scaling obtained in approximate calculations of flux noise in the local regime in thin film wires \cite{Bialczak2007, Laforest2015} ($W$ is the lateral width of the thin film). 

The flux noise power in the bulk is given by
\begin{eqnarray}
\langle \left(\delta\Phi\right)^{2}\rangle &=& \frac{S(S+1)}{3} \sigma_3 (g\mu_B)^2 \int_{{\rm SC}}d^3 r \left|\bm{F}(\bm{r})\right|^{2}\nonumber\\
&=& \frac{S(S+1)}{3} \sigma_3 (g\mu_B)^2 \mu_0 L_{{\rm int}}.
\end{eqnarray}
Note how the bulk noise power is directly proportional to the internal inductance $L_{{\rm int}}$, which is sensitive to nonlocal effects. 

\begin{figure}[ht]
    \includegraphics[width = 0.49\textwidth]{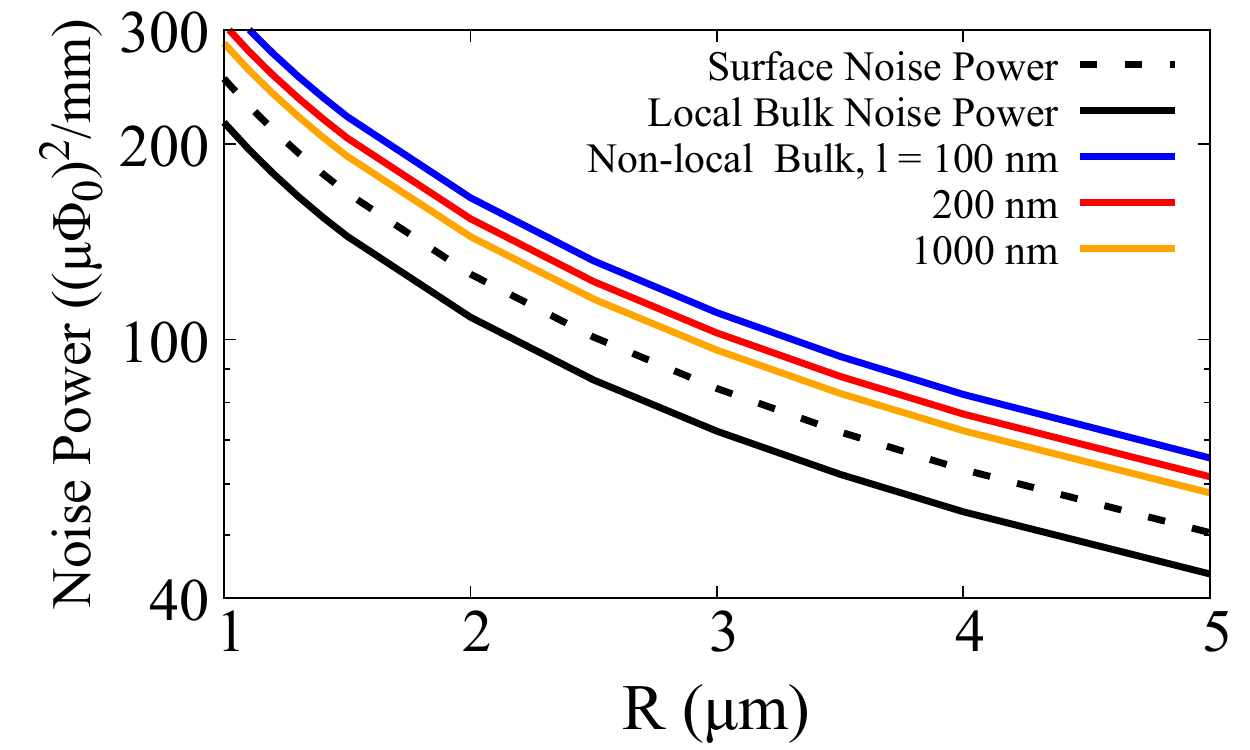}
    \caption[Flux Noise Power Infinite Cylinder: 2D Pippard]{Calculated flux noise power due to the presence of spin impurities in the cylindrical wire, for the local and nonlocal cases with $\lambda,\xi$ as in Fig.~\ref{multixi}. 
    We separate flux noise into contributions from spins at the surface (spin density $\sigma_2=5\times 10^{12}$~cm$^{-2}$) and in the bulk (spin density $\sigma_3=1.3\times 10^{18}$~cm$^{-3}$).}
    \label{fig:flux_noise_power}
\end{figure}

Figure~\ref{fig:flux_noise_power} shows calculations of the flux noise power due to surface and bulk spins as a function of wire radius, for $\lambda,\xi$ as in Fig.~\ref{multixi}.
We assumed $S=1/2$ impurities with surface density $\sigma_2=5\times 10^{12}$~cm$^{-2}$ and bulk density $\sigma_3=1.3\times 10^{18}$~cm$^{-3}$ (both $\sigma_2$ and $\sigma_3$ correspond to average distance between impurities $\sim 1$~$\mu$m).
Here we see that the bulk flux noise power increases significantly as $l$ increases. For  the local case the surface noise power is larger than the bulk one; in contrast for the nonlocal case the 
bulk noise power is 30\% larger than the surface one, giving the dominant contribution to flux noise.

\section{Conclusions\label{conclusions}}

In summary, we presented an exact numerical method to calculate the space dependence of the vector potential $\bm{A}$ and current density $\bm{J}$ in nonlocal superconductors using exact integration of the Fredholm integral equations. We showed numerical results for a cylindrical wire, and applied them to calculations of inductance and flux noise. 

In the presence of nonlocal effects we showed that both $\bm{A}$ and $\bm{J}$ change sign inside the wire, due to \emph{self-induced overscreening}.
To our knowledge this is the first time that overscreening without an external magnetic field is reported. Apart from the overscreening, the vector potential was shown 
to be reasonably described by the local theory with a renormalized penetration depth $\lambda_{R} = 0.85 (\lambda^{2}\xi)^{\frac{1}{3}}$. It is interesting to note that this choice of $\lambda_R$ is specific to the cylindrical surface and is different from the result obtained in a flat surface (For a flat surface, $\lambda_{R} = 0.65 (\lambda^2\xi_0)^{\frac{1}{3}}$, see Appendix 3 in \cite{Tinkham1996}). 

In the presence of nonlocality the London relation $\bm{J}=-\bm{A}/(\mu_0 \lambda^{2})$ ceases to hold, in that $\bm{J}(\bm{r})$ develops a peak away from the surface of the wire. We also find that $\bm{J}$ decreases slower than $\bm{A}$ towards the centre of the wire. We compared our exact results to the usual approximation of evaluating $\bm{J}$ from the Pippard relation with the $\lambda_R$ approximation of $\bm{A}$ as an input. 
Our Fig.~\ref{figJmod} showed that this approximation is reliable only in the region near the surface of the wire.

We showed that nonlocal electrodynamics has a large impact on device properties such as inductance and flux noise. While for local superconductors kinetic inductance is always larger than internal inductance, the oposite happens for nonlocal superconductors. 

We also considered the impact of nonlocality on the flux noise produced by the time-dependent fluctuation of impurities with spin. The amount of noise caused by each impurity scales quadratically with the magnetic field at the impurity location, see Eqs.~(\ref{phi_f_s})~and~(\ref{fi}). This observation has led many authors to assume that spins at the surface are the only ones causing flux noise \cite{Koch2007, deSousa2007, Bialczak2007, Lanting2014, Laforest2015}. The fact that the magnetic field at the surface of a cylindrical wire only depends on the total current inside the wire implies that the flux noise due to impurities at the surface is independent of nonlocality. This is in stark contrast to the flux noise due to impurities in the bulk. As $\xi$ increases, more magnetic field penetrates inside the wire, making the impurities in the bulk more relevant. 
As a result, devices made of nonlocal superconductors can be more sensitive to spin impurities in the bulk rather than in the surface. It all depends on the relative sizes of the surface and bulk impurity densities, and of $\xi$. 

In conclusion, we demonstrated that nonlocal electrodynamics has large impact on the circuit properties of superconducting wires. We hope that our results will contribute to modelling and design of superconducting devices with Pippard cherence length comparable to wire length scales, as is often the case for devices made of aluminum. 

\begin{acknowledgments}
We acknowledge financial support from NSERC (Canada) through its Discovery (RGPIN-2015- 03938) and Collaborative Research and Development programs (CRDPJ 478366-14). We thank M. Amin, I. Diniz, R. Harris, T. Lanting, and M. Le Dall for useful discussions. We are indebted to W. Guichard for critical reading of the manuscript and for providing numerous suggestions.
\end{acknowledgments}

\bibliography{reference}

\begin{thebibliography}{26}%
\makeatletter
\providecommand \@ifxundefined [1]{%
 \@ifx{#1\undefined}
}%
\providecommand \@ifnum [1]{%
 \ifnum #1\expandafter \@firstoftwo
 \else \expandafter \@secondoftwo
 \fi
}%
\providecommand \@ifx [1]{%
 \ifx #1\expandafter \@firstoftwo
 \else \expandafter \@secondoftwo
 \fi
}%
\providecommand \natexlab [1]{#1}%
\providecommand \enquote  [1]{``#1''}%
\providecommand \bibnamefont  [1]{#1}%
\providecommand \bibfnamefont [1]{#1}%
\providecommand \citenamefont [1]{#1}%
\providecommand \href@noop [0]{\@secondoftwo}%
\providecommand \href [0]{\begingroup \@sanitize@url \@href}%
\providecommand \@href[1]{\@@startlink{#1}\@@href}%
\providecommand \@@href[1]{\endgroup#1\@@endlink}%
\providecommand \@sanitize@url [0]{\catcode `\\12\catcode `\$12\catcode
  `\&12\catcode `\#12\catcode `\^12\catcode `\_12\catcode `\%12\relax}%
\providecommand \@@startlink[1]{}%
\providecommand \@@endlink[0]{}%
\providecommand \url  [0]{\begingroup\@sanitize@url \@url }%
\providecommand \@url [1]{\endgroup\@href {#1}{\urlprefix }}%
\providecommand \urlprefix  [0]{URL }%
\providecommand \Eprint [0]{\href }%
\providecommand \doibase [0]{http://dx.doi.org/}%
\providecommand \selectlanguage [0]{\@gobble}%
\providecommand \bibinfo  [0]{\@secondoftwo}%
\providecommand \bibfield  [0]{\@secondoftwo}%
\providecommand \translation [1]{[#1]}%
\providecommand \BibitemOpen [0]{}%
\providecommand \bibitemStop [0]{}%
\providecommand \bibitemNoStop [0]{.\EOS\space}%
\providecommand \EOS [0]{\spacefactor3000\relax}%
\providecommand \BibitemShut  [1]{\csname bibitem#1\endcsname}%
\let\auto@bib@innerbib\@empty
\bibitem [{\citenamefont {Clarke}\ and\ \citenamefont
  {Wilhelm}(2008)}]{Clarke2008}%
  \BibitemOpen
  \bibfield  {author} {\bibinfo {author} {\bibfnamefont {John}\ \bibnamefont
  {Clarke}}\ and\ \bibinfo {author} {\bibfnamefont {Frank~K.}\ \bibnamefont
  {Wilhelm}},\ }\bibfield  {title} {\enquote {\bibinfo {title}
  {{Superconducting quantum bits}},}\ }\href {\doibase 10.1038/nature07128}
  {\bibfield  {journal} {\bibinfo  {journal} {Nature}\ }\textbf {\bibinfo
  {volume} {453}},\ \bibinfo {pages} {1031} (\bibinfo {year}
  {2008})}\BibitemShut {NoStop}%
\bibitem [{\citenamefont {Denchev}\ \emph {et~al.}(2016)\citenamefont
  {Denchev}, \citenamefont {Boixo}, \citenamefont {Isakov}, \citenamefont
  {Ding}, \citenamefont {Babbush}, \citenamefont {Smelyanskiy}, \citenamefont
  {Martinis},\ and\ \citenamefont {Neven}}]{Denchev2016}%
  \BibitemOpen
  \bibfield  {author} {\bibinfo {author} {\bibfnamefont {Vasil~S.}\
  \bibnamefont {Denchev}}, \bibinfo {author} {\bibfnamefont {Sergio}\
  \bibnamefont {Boixo}}, \bibinfo {author} {\bibfnamefont {Sergei~V.}\
  \bibnamefont {Isakov}}, \bibinfo {author} {\bibfnamefont {Nan}\ \bibnamefont
  {Ding}}, \bibinfo {author} {\bibfnamefont {Ryan}\ \bibnamefont {Babbush}},
  \bibinfo {author} {\bibfnamefont {Vadim}\ \bibnamefont {Smelyanskiy}},
  \bibinfo {author} {\bibfnamefont {John}\ \bibnamefont {Martinis}}, \ and\
  \bibinfo {author} {\bibfnamefont {Hartmut}\ \bibnamefont {Neven}},\
  }\bibfield  {title} {\enquote {\bibinfo {title} {{What is the computational
  value of finite-range tunneling?}}}\ }\href {\doibase
  10.1103/PhysRevX.6.031015} {\bibfield  {journal} {\bibinfo  {journal} {Phys.
  Rev. X}\ }\textbf {\bibinfo {volume} {6}},\ \bibinfo {pages} {031015}
  (\bibinfo {year} {2016})}\BibitemShut {NoStop}%
\bibitem [{\citenamefont {Barends}\ \emph {et~al.}(2016)\citenamefont
  {Barends}, \citenamefont {Shabani}, \citenamefont {Lamata}, \citenamefont
  {Kelly}, \citenamefont {Mezzacapo}, \citenamefont {Heras}, \citenamefont
  {Babbush}, \citenamefont {Fowler}, \citenamefont {Campbell}, \citenamefont
  {Chen}, \citenamefont {Chen}, \citenamefont {Chiaro}, \citenamefont
  {Dunsworth}, \citenamefont {Jeffrey}, \citenamefont {Lucero}, \citenamefont
  {Megrant}, \citenamefont {Mutus}, \citenamefont {Neeley}, \citenamefont
  {Neill}, \citenamefont {O'Malley}, \citenamefont {Quintana}, \citenamefont
  {Roushan}, \citenamefont {Sank}, \citenamefont {Vainsencher}, \citenamefont
  {Wenner}, \citenamefont {White}, \citenamefont {Solano}, \citenamefont
  {Neven},\ and\ \citenamefont {Martinis}}]{Barends2016}%
  \BibitemOpen
  \bibfield  {author} {\bibinfo {author} {\bibfnamefont {R.}~\bibnamefont
  {Barends}}, \bibinfo {author} {\bibfnamefont {A.}~\bibnamefont {Shabani}},
  \bibinfo {author} {\bibfnamefont {L.}~\bibnamefont {Lamata}}, \bibinfo
  {author} {\bibfnamefont {J.}~\bibnamefont {Kelly}}, \bibinfo {author}
  {\bibfnamefont {A.}~\bibnamefont {Mezzacapo}}, \bibinfo {author}
  {\bibfnamefont {U.~Las}\ \bibnamefont {Heras}}, \bibinfo {author}
  {\bibfnamefont {R.}~\bibnamefont {Babbush}}, \bibinfo {author} {\bibfnamefont
  {A.~G.}\ \bibnamefont {Fowler}}, \bibinfo {author} {\bibfnamefont
  {B.}~\bibnamefont {Campbell}}, \bibinfo {author} {\bibfnamefont
  {Yu}~\bibnamefont {Chen}}, \bibinfo {author} {\bibfnamefont {Z.}~\bibnamefont
  {Chen}}, \bibinfo {author} {\bibfnamefont {B.}~\bibnamefont {Chiaro}},
  \bibinfo {author} {\bibfnamefont {A.}~\bibnamefont {Dunsworth}}, \bibinfo
  {author} {\bibfnamefont {E.}~\bibnamefont {Jeffrey}}, \bibinfo {author}
  {\bibfnamefont {E.}~\bibnamefont {Lucero}}, \bibinfo {author} {\bibfnamefont
  {A.}~\bibnamefont {Megrant}}, \bibinfo {author} {\bibfnamefont {J.~Y.}\
  \bibnamefont {Mutus}}, \bibinfo {author} {\bibfnamefont {M.}~\bibnamefont
  {Neeley}}, \bibinfo {author} {\bibfnamefont {C.}~\bibnamefont {Neill}},
  \bibinfo {author} {\bibfnamefont {P.~J.~J.}\ \bibnamefont {O'Malley}},
  \bibinfo {author} {\bibfnamefont {C.}~\bibnamefont {Quintana}}, \bibinfo
  {author} {\bibfnamefont {P.}~\bibnamefont {Roushan}}, \bibinfo {author}
  {\bibfnamefont {D.}~\bibnamefont {Sank}}, \bibinfo {author} {\bibfnamefont
  {A.}~\bibnamefont {Vainsencher}}, \bibinfo {author} {\bibfnamefont
  {J.}~\bibnamefont {Wenner}}, \bibinfo {author} {\bibfnamefont {T.~C.}\
  \bibnamefont {White}}, \bibinfo {author} {\bibfnamefont {E.}~\bibnamefont
  {Solano}}, \bibinfo {author} {\bibfnamefont {H.}~\bibnamefont {Neven}}, \
  and\ \bibinfo {author} {\bibfnamefont {John~M.}\ \bibnamefont {Martinis}},\
  }\bibfield  {title} {\enquote {\bibinfo {title} {{Digitized adiabatic quantum
  computing with a superconducting circuit}},}\ }\href {\doibase
  10.1038/nature17658} {\bibfield  {journal} {\bibinfo  {journal} {Nature}\
  }\textbf {\bibinfo {volume} {534}},\ \bibinfo {pages} {222} (\bibinfo {year}
  {2016})}\BibitemShut {NoStop}%
\bibitem [{\citenamefont {Fourie}(2017)}]{InductEx2017}%
  \BibitemOpen
  \bibfield  {author} {\bibinfo {author} {\bibfnamefont {C.~J.}\ \bibnamefont
  {Fourie}},\ }\href@noop {} {\emph {\bibinfo {title} {InductEx User's Guide
  Version 5.07}}}\ (\bibinfo  {publisher} {Stellenbosch University},\ \bibinfo
  {year} {2017})\BibitemShut {NoStop}%
\bibitem [{\citenamefont {Pippard}(1953)}]{Pippard1953}%
  \BibitemOpen
  \bibfield  {author} {\bibinfo {author} {\bibfnamefont {A.~B.}\ \bibnamefont
  {Pippard}},\ }\bibfield  {title} {\enquote {\bibinfo {title} {{An
  experimental and theoretical study of the relation between magnetic field and
  current in a superconductor}},}\ }\href {\doibase 10.1098/rspa.1953.0040}
  {\bibfield  {journal} {\bibinfo  {journal} {Proc. Roy. Soc. (London)}\
  }\textbf {\bibinfo {volume} {A216}},\ \bibinfo {pages} {547} (\bibinfo {year}
  {1953})}\BibitemShut {NoStop}%
\bibitem [{\citenamefont {Belzig}(1999)}]{Belzig1999}%
  \BibitemOpen
  \bibfield  {author} {\bibinfo {author} {\bibfnamefont {Wolfgang}\
  \bibnamefont {Belzig}},\ }\emph {\bibinfo {title} {{Magnetic and Spectral
  Properties of Superconducting Proximity Systems}}},\ \href@noop {} {Ph.D.
  thesis},\ \bibinfo  {school} {University of Karlsruhe} (\bibinfo {year}
  {1999})\BibitemShut {NoStop}%
\bibitem [{\citenamefont {Tinkham}(1996)}]{Tinkham1996}%
  \BibitemOpen
  \bibfield  {author} {\bibinfo {author} {\bibfnamefont {M.}~\bibnamefont
  {Tinkham}},\ }\href@noop {} {\emph {\bibinfo {title} {Introduction to
  Superconductivity}}},\ \bibinfo {edition} {2nd}\ ed.\ (\bibinfo  {publisher}
  {McGraw-Hill, New York},\ \bibinfo {year} {1996})\BibitemShut {NoStop}%
\bibitem [{\citenamefont {Rhoderick}\ and\ \citenamefont
  {Wilson}(1962)}]{Rhoderick1962}%
  \BibitemOpen
  \bibfield  {author} {\bibinfo {author} {\bibfnamefont {E.~H.}\ \bibnamefont
  {Rhoderick}}\ and\ \bibinfo {author} {\bibfnamefont {E.~M.}\ \bibnamefont
  {Wilson}},\ }\bibfield  {title} {\enquote {\bibinfo {title} {{Current
  Distribution in Thin Superconducting Films}},}\ }\href {\doibase
  10.1038/1941167b0} {\bibfield  {journal} {\bibinfo  {journal} {Nature}\
  }\textbf {\bibinfo {volume} {194}},\ \bibinfo {pages} {1167} (\bibinfo {year}
  {1962})}\BibitemShut {NoStop}%
\bibitem [{\citenamefont {Lee}\ \emph {et~al.}(1994)\citenamefont {Lee},
  \citenamefont {Orlando},\ and\ \citenamefont {Lyons}}]{Lee1994}%
  \BibitemOpen
  \bibfield  {author} {\bibinfo {author} {\bibfnamefont {Laurence~H.}\
  \bibnamefont {Lee}}, \bibinfo {author} {\bibfnamefont {Terry~P.}\
  \bibnamefont {Orlando}}, \ and\ \bibinfo {author} {\bibfnamefont
  {W.~Gregory}\ \bibnamefont {Lyons}},\ }\bibfield  {title} {\enquote {\bibinfo
  {title} {{Current Distribution in Superconducting Thin-Film Strips}},}\
  }\href {\doibase 10.1109/77.273063} {\bibfield  {journal} {\bibinfo
  {journal} {IEEE Trans. Appl. Supercond.}\ }\textbf {\bibinfo {volume} {4}},\
  \bibinfo {pages} {41} (\bibinfo {year} {1994})}\BibitemShut {NoStop}%
\bibitem [{\citenamefont {Anton}\ \emph {et~al.}(2013)\citenamefont {Anton},
  \citenamefont {Sognnaes}, \citenamefont {Birenbaum}, \citenamefont
  {O'Kelley}, \citenamefont {Fourie},\ and\ \citenamefont
  {Clarke}}]{Anton2013b}%
  \BibitemOpen
  \bibfield  {author} {\bibinfo {author} {\bibfnamefont {S.~M.}\ \bibnamefont
  {Anton}}, \bibinfo {author} {\bibfnamefont {I.~A.~B.}\ \bibnamefont
  {Sognnaes}}, \bibinfo {author} {\bibfnamefont {J.~S.}\ \bibnamefont
  {Birenbaum}}, \bibinfo {author} {\bibfnamefont {S.~R.}\ \bibnamefont
  {O'Kelley}}, \bibinfo {author} {\bibfnamefont {C.~J.}\ \bibnamefont
  {Fourie}}, \ and\ \bibinfo {author} {\bibfnamefont {J.}~\bibnamefont
  {Clarke}},\ }\bibfield  {title} {\enquote {\bibinfo {title} {{Mean square
  flux noise in SQUIDs and qubits: numerical calculations}},}\ }\href {\doibase
  10.1088/0953-2048/26/7/075022} {\bibfield  {journal} {\bibinfo  {journal}
  {Supercond. Sci. Technol.}\ }\textbf {\bibinfo {volume} {26}},\ \bibinfo
  {pages} {075022} (\bibinfo {year} {2013})}\BibitemShut {NoStop}%
\bibitem [{\citenamefont {Van~Duzer}\ and\ \citenamefont
  {Turner}(1999)}]{VanDuzer1999}%
  \BibitemOpen
  \bibfield  {author} {\bibinfo {author} {\bibfnamefont {T.}~\bibnamefont
  {Van~Duzer}}\ and\ \bibinfo {author} {\bibfnamefont {C.~W.}\ \bibnamefont
  {Turner}},\ }\href@noop {} {\emph {\bibinfo {title} {Superconductive Devices
  and Circuits}}},\ \bibinfo {edition} {2nd}\ ed.\ (\bibinfo  {publisher}
  {Prentice Hall},\ \bibinfo {year} {1999})\BibitemShut {NoStop}%
\bibitem [{\citenamefont {Cooper}(1961)}]{Cooper1961}%
  \BibitemOpen
  \bibfield  {author} {\bibinfo {author} {\bibfnamefont {L.~N.}\ \bibnamefont
  {Cooper}},\ }\bibfield  {title} {\enquote {\bibinfo {title} {Current flow in
  thin superconducting films},}\ }in\ \href@noop {} {\emph {\bibinfo
  {booktitle} {Proc. VII Int. Conf. Low Temp. Phys.}}}\ (\bibinfo {year}
  {1961})\ p.\ \bibinfo {pages} {416}\BibitemShut {NoStop}%
\bibitem [{\citenamefont {Marcus}(1961)}]{Marcus1961}%
  \BibitemOpen
  \bibfield  {author} {\bibinfo {author} {\bibfnamefont {P.~M.}\ \bibnamefont
  {Marcus}},\ }\bibfield  {title} {\enquote {\bibinfo {title} {Currents and
  fields in a superconducting film carrying a steady current},}\ }in\
  \href@noop {} {\emph {\bibinfo {booktitle} {Proc. VII Int. Conf. Low Temp.
  Phys.}}}\ (\bibinfo {year} {1961})\ p.\ \bibinfo {pages} {418}\BibitemShut
  {NoStop}%
\bibitem [{\citenamefont {LaForest}\ and\ \citenamefont
  {de~Sousa}(2015)}]{Laforest2015}%
  \BibitemOpen
  \bibfield  {author} {\bibinfo {author} {\bibfnamefont {S.}~\bibnamefont
  {LaForest}}\ and\ \bibinfo {author} {\bibfnamefont {R.}~\bibnamefont
  {de~Sousa}},\ }\bibfield  {title} {\enquote {\bibinfo {title} {Flux-vector
  model of spin noise in superconducting circuits: Electron versus nuclear
  spins and role of phase transition},}\ }\href {\doibase
  10.1103/PhysRevB.92.054502} {\bibfield  {journal} {\bibinfo  {journal} {Phys.
  Rev. B}\ }\textbf {\bibinfo {volume} {92}},\ \bibinfo {pages} {054502}
  (\bibinfo {year} {2015})}\BibitemShut {NoStop}%
\bibitem [{\citenamefont {Trim}(1990)}]{Trim1990}%
  \BibitemOpen
  \bibfield  {author} {\bibinfo {author} {\bibfnamefont {D.~W.}\ \bibnamefont
  {Trim}},\ }\href@noop {} {\emph {\bibinfo {title} {Applied Partial
  Differential Equations}}}\ (\bibinfo  {publisher} {PWS-Kent},\ \bibinfo
  {year} {1990})\BibitemShut {NoStop}%
\bibitem [{\citenamefont {Atkinson}\ and\ \citenamefont
  {Shampine}(2008)}]{Atkinson2007}%
  \BibitemOpen
  \bibfield  {author} {\bibinfo {author} {\bibfnamefont {K.~E.}\ \bibnamefont
  {Atkinson}}\ and\ \bibinfo {author} {\bibfnamefont {L.~F.}\ \bibnamefont
  {Shampine}},\ }\bibfield  {title} {\enquote {\bibinfo {title} {Algorithm 876:
  Solving fredholm integral equations of the second kind in matlab},}\ }\href
  {\doibase 10.1145/1377596.1377601} {\bibfield  {journal} {\bibinfo  {journal}
  {ACM Trans. Math. Softw.}\ }\textbf {\bibinfo {volume} {34}},\ \bibinfo
  {pages} {21} (\bibinfo {year} {2008})}\BibitemShut {NoStop}%
\bibitem [{\citenamefont {Drangeid}\ and\ \citenamefont
  {Sommerhalder}(1962)}]{Drangeid1962}%
  \BibitemOpen
  \bibfield  {author} {\bibinfo {author} {\bibfnamefont {K.~E.}\ \bibnamefont
  {Drangeid}}\ and\ \bibinfo {author} {\bibfnamefont {R.}~\bibnamefont
  {Sommerhalder}},\ }\bibfield  {title} {\enquote {\bibinfo {title} {{Observed
  sign reversal of a magnetic field penetrating a superconductor}},}\ }\href
  {\doibase 10.1103/PhysRevLett.8.467} {\bibfield  {journal} {\bibinfo
  {journal} {Phys. Rev. Lett.}\ }\textbf {\bibinfo {volume} {8}},\ \bibinfo
  {pages} {467} (\bibinfo {year} {1962})}\BibitemShut {NoStop}%
\bibitem [{\citenamefont {Mazin}\ \emph {et~al.}(2002)\citenamefont {Mazin},
  \citenamefont {Day}, \citenamefont {LeDuc}, \citenamefont {Vayonakis},\ and\
  \citenamefont {Zmuidzinas}}]{Mazin2002}%
  \BibitemOpen
  \bibfield  {author} {\bibinfo {author} {\bibfnamefont {B.~A.}\ \bibnamefont
  {Mazin}}, \bibinfo {author} {\bibfnamefont {P.~K.}\ \bibnamefont {Day}},
  \bibinfo {author} {\bibfnamefont {H.~G.}\ \bibnamefont {LeDuc}}, \bibinfo
  {author} {\bibfnamefont {A.}~\bibnamefont {Vayonakis}}, \ and\ \bibinfo
  {author} {\bibfnamefont {J.}~\bibnamefont {Zmuidzinas}},\ }\bibfield  {title}
  {\enquote {\bibinfo {title} {{Superconducting kinetic inductance photon
  detectors}},}\ }\href {\doibase 10.1117/12.460456} {\bibfield  {journal}
  {\bibinfo  {journal} {Proc. SPIE}\ }\textbf {\bibinfo {volume} {4849}},\
  \bibinfo {pages} {283} (\bibinfo {year} {2002})}\BibitemShut {NoStop}%
\bibitem [{\citenamefont {Koch}\ \emph {et~al.}(1983)\citenamefont {Koch},
  \citenamefont {Clarke}, \citenamefont {Goubau}, \citenamefont {Martinis},
  \citenamefont {Pegrum},\ and\ \citenamefont {Harlingen}}]{Koch1983}%
  \BibitemOpen
  \bibfield  {author} {\bibinfo {author} {\bibfnamefont {R.~H.}\ \bibnamefont
  {Koch}}, \bibinfo {author} {\bibfnamefont {J.}~\bibnamefont {Clarke}},
  \bibinfo {author} {\bibfnamefont {W.~M.}\ \bibnamefont {Goubau}}, \bibinfo
  {author} {\bibfnamefont {J.~M.}\ \bibnamefont {Martinis}}, \bibinfo {author}
  {\bibfnamefont {C.~M.}\ \bibnamefont {Pegrum}}, \ and\ \bibinfo {author}
  {\bibfnamefont {D.~J.}\ \bibnamefont {Harlingen}},\ }\bibfield  {title}
  {\enquote {\bibinfo {title} {{Flicker (1/f) noise in tunnel junction dc
  SQUIDS}},}\ }\href {\doibase 10.1007/BF00683423} {\bibfield  {journal}
  {\bibinfo  {journal} {J. Low Temp. Phys.}\ }\textbf {\bibinfo {volume}
  {51}},\ \bibinfo {pages} {207} (\bibinfo {year} {1983})}\BibitemShut
  {NoStop}%
\bibitem [{\citenamefont {Wellstood}\ \emph {et~al.}(1987)\citenamefont
  {Wellstood}, \citenamefont {Urbina},\ and\ \citenamefont
  {Clarke}}]{Wellstood1987}%
  \BibitemOpen
  \bibfield  {author} {\bibinfo {author} {\bibfnamefont {F.~C.}\ \bibnamefont
  {Wellstood}}, \bibinfo {author} {\bibfnamefont {C.}~\bibnamefont {Urbina}}, \
  and\ \bibinfo {author} {\bibfnamefont {J.}~\bibnamefont {Clarke}},\
  }\bibfield  {title} {\enquote {\bibinfo {title} {{Low frequency noise in dc
  superconducting quantum interference devices below 1 K}},}\ }\href {\doibase
  10.1063/1.98041} {\bibfield  {journal} {\bibinfo  {journal} {Appl. Phys.
  Lett.}\ }\textbf {\bibinfo {volume} {50}},\ \bibinfo {pages} {772} (\bibinfo
  {year} {1987})}\BibitemShut {NoStop}%
\bibitem [{\citenamefont {Koch}\ \emph {et~al.}(2007)\citenamefont {Koch},
  \citenamefont {DiVincenzo},\ and\ \citenamefont {Clarke}}]{Koch2007}%
  \BibitemOpen
  \bibfield  {author} {\bibinfo {author} {\bibfnamefont {R.~H.}\ \bibnamefont
  {Koch}}, \bibinfo {author} {\bibfnamefont {D.~P.}\ \bibnamefont
  {DiVincenzo}}, \ and\ \bibinfo {author} {\bibfnamefont {J.}~\bibnamefont
  {Clarke}},\ }\bibfield  {title} {\enquote {\bibinfo {title} {{Model for 1/f
  Flux Noise in SQUIDs and Qubits}},}\ }\href {\doibase
  10.1103/PhysRevLett.98.267003} {\bibfield  {journal} {\bibinfo  {journal}
  {Phys. Rev. Lett.}\ }\textbf {\bibinfo {volume} {98}},\ \bibinfo {pages}
  {267003} (\bibinfo {year} {2007})}\BibitemShut {NoStop}%
\bibitem [{\citenamefont {de~Sousa}(2007)}]{deSousa2007}%
  \BibitemOpen
  \bibfield  {author} {\bibinfo {author} {\bibfnamefont {R.}~\bibnamefont
  {de~Sousa}},\ }\bibfield  {title} {\enquote {\bibinfo {title} {{Dangling-bond
  spin relaxation and magnetic 1/f noise from the amorphous-semiconductor/oxide
  interface: Theory}},}\ }\href {\doibase 10.1103/PhysRevB.76.245306}
  {\bibfield  {journal} {\bibinfo  {journal} {Phys. Rev. B}\ }\textbf {\bibinfo
  {volume} {76}},\ \bibinfo {pages} {245306} (\bibinfo {year}
  {2007})}\BibitemShut {NoStop}%
\bibitem [{\citenamefont {Sendelbach}\ \emph {et~al.}(2008)\citenamefont
  {Sendelbach}, \citenamefont {Hover}, \citenamefont {Kittel}, \citenamefont
  {M{\"{u}}ck}, \citenamefont {Martinis},\ and\ \citenamefont
  {McDermott}}]{Sendelbach2008}%
  \BibitemOpen
  \bibfield  {author} {\bibinfo {author} {\bibfnamefont {S.}~\bibnamefont
  {Sendelbach}}, \bibinfo {author} {\bibfnamefont {D.}~\bibnamefont {Hover}},
  \bibinfo {author} {\bibfnamefont {A.}~\bibnamefont {Kittel}}, \bibinfo
  {author} {\bibfnamefont {M.}~\bibnamefont {M{\"{u}}ck}}, \bibinfo {author}
  {\bibfnamefont {J.~M.}\ \bibnamefont {Martinis}}, \ and\ \bibinfo {author}
  {\bibfnamefont {R.}~\bibnamefont {McDermott}},\ }\bibfield  {title} {\enquote
  {\bibinfo {title} {{Magnetism in SQUIDs at Millikelvin Temperatures}},}\
  }\href {\doibase 10.1103/PhysRevLett.100.227006} {\bibfield  {journal}
  {\bibinfo  {journal} {Phys. Rev. Lett.}\ }\textbf {\bibinfo {volume} {100}},\
  \bibinfo {pages} {227006} (\bibinfo {year} {2008})}\BibitemShut {NoStop}%
\bibitem [{\citenamefont {Lanting}\ \emph {et~al.}(2014)\citenamefont
  {Lanting}, \citenamefont {Amin}, \citenamefont {Berkley}, \citenamefont
  {Rich}, \citenamefont {Chen}, \citenamefont {LaForest},\ and\ \citenamefont
  {de~Sousa}}]{Lanting2014}%
  \BibitemOpen
  \bibfield  {author} {\bibinfo {author} {\bibfnamefont {T.}~\bibnamefont
  {Lanting}}, \bibinfo {author} {\bibfnamefont {M.~H.}\ \bibnamefont {Amin}},
  \bibinfo {author} {\bibfnamefont {A.~J.}\ \bibnamefont {Berkley}}, \bibinfo
  {author} {\bibfnamefont {C.}~\bibnamefont {Rich}}, \bibinfo {author}
  {\bibfnamefont {S.-F.}\ \bibnamefont {Chen}}, \bibinfo {author}
  {\bibfnamefont {S.}~\bibnamefont {LaForest}}, \ and\ \bibinfo {author}
  {\bibfnamefont {Rog{\'{e}}rio}\ \bibnamefont {de~Sousa}},\ }\bibfield
  {title} {\enquote {\bibinfo {title} {{Evidence for temperature-dependent spin
  diffusion as a mechanism of intrinsic flux noise in SQUIDs}},}\ }\href
  {\doibase 10.1103/PhysRevB.89.014503} {\bibfield  {journal} {\bibinfo
  {journal} {Phys. Rev. B}\ }\textbf {\bibinfo {volume} {89}},\ \bibinfo
  {pages} {014503} (\bibinfo {year} {2014})}\BibitemShut {NoStop}%
\bibitem [{\citenamefont {Kumar}\ \emph {et~al.}(2016)\citenamefont {Kumar},
  \citenamefont {Sendelbach}, \citenamefont {Beck}, \citenamefont {Freeland},
  \citenamefont {Wang}, \citenamefont {Wang}, \citenamefont {Yu}, \citenamefont
  {Wu}, \citenamefont {Pappas},\ and\ \citenamefont {McDermott}}]{Kumar2016}%
  \BibitemOpen
  \bibfield  {author} {\bibinfo {author} {\bibfnamefont {P.}~\bibnamefont
  {Kumar}}, \bibinfo {author} {\bibfnamefont {S.}~\bibnamefont {Sendelbach}},
  \bibinfo {author} {\bibfnamefont {M.~A.}\ \bibnamefont {Beck}}, \bibinfo
  {author} {\bibfnamefont {J.~W.}\ \bibnamefont {Freeland}}, \bibinfo {author}
  {\bibfnamefont {Zhe}\ \bibnamefont {Wang}}, \bibinfo {author} {\bibfnamefont
  {Hui}\ \bibnamefont {Wang}}, \bibinfo {author} {\bibfnamefont {Clare~C.}\
  \bibnamefont {Yu}}, \bibinfo {author} {\bibfnamefont {R.~Q.}\ \bibnamefont
  {Wu}}, \bibinfo {author} {\bibfnamefont {D.~P.}\ \bibnamefont {Pappas}}, \
  and\ \bibinfo {author} {\bibfnamefont {R.}~\bibnamefont {McDermott}},\
  }\bibfield  {title} {\enquote {\bibinfo {title} {{Origin and Reduction of 1/f
  Magnetic Flux Noise in Superconducting Devices}},}\ }\href {\doibase
  10.1103/PhysRevApplied.6.041001} {\bibfield  {journal} {\bibinfo  {journal}
  {Phys. Rev. Appl.}\ }\textbf {\bibinfo {volume} {6}},\ \bibinfo {pages}
  {041001} (\bibinfo {year} {2016})}\BibitemShut {NoStop}%
\bibitem [{\citenamefont {Bialczak}\ \emph {et~al.}(2007)\citenamefont
  {Bialczak}, \citenamefont {McDermott}, \citenamefont {Ansmann}, \citenamefont
  {Hofheinz}, \citenamefont {Katz}, \citenamefont {Lucero}, \citenamefont
  {Neeley}, \citenamefont {O'Connell}, \citenamefont {Wang}, \citenamefont
  {Cleland},\ and\ \citenamefont {Martinis}}]{Bialczak2007}%
  \BibitemOpen
  \bibfield  {author} {\bibinfo {author} {\bibfnamefont {R.~C.}\ \bibnamefont
  {Bialczak}}, \bibinfo {author} {\bibfnamefont {R.}~\bibnamefont {McDermott}},
  \bibinfo {author} {\bibfnamefont {M.}~\bibnamefont {Ansmann}}, \bibinfo
  {author} {\bibfnamefont {M.}~\bibnamefont {Hofheinz}}, \bibinfo {author}
  {\bibfnamefont {N.}~\bibnamefont {Katz}}, \bibinfo {author} {\bibfnamefont
  {Erik}\ \bibnamefont {Lucero}}, \bibinfo {author} {\bibfnamefont {Matthew}\
  \bibnamefont {Neeley}}, \bibinfo {author} {\bibfnamefont {A.~D.}\
  \bibnamefont {O'Connell}}, \bibinfo {author} {\bibfnamefont {H.}~\bibnamefont
  {Wang}}, \bibinfo {author} {\bibfnamefont {A.~N.}\ \bibnamefont {Cleland}}, \
  and\ \bibinfo {author} {\bibfnamefont {John~M.}\ \bibnamefont {Martinis}},\
  }\bibfield  {title} {\enquote {\bibinfo {title} {{1/f Flux Noise in Josephson
  Phase Qubits}},}\ }\href {\doibase 10.1103/PhysRevLett.99.187006} {\bibfield
  {journal} {\bibinfo  {journal} {Phys. Rev. Lett.}\ }\textbf {\bibinfo
  {volume} {99}},\ \bibinfo {pages} {187006} (\bibinfo {year}
  {2007})}\BibitemShut {NoStop}%
\end{thebibliography}%

\end{document}